\documentclass[journal=jacsat,manuscript=article]{achemso}

\usepackage{multirow}
\usepackage{makecell}
\usepackage{cellspace}
\usepackage{booktabs}
\usepackage{amsmath} 
\usepackage{amssymb} 
\usepackage{dblfloatfix}
\usepackage[section]{placeins}

\title{Graph-Based Kirchhoff Modeling of Non-Ohmic Electron Transport in Self-Assembled Nanonecklace Networks}

\author{Obed Issakah}
\affiliation[cheme]
{Department of Chemical and Biomolecular Engineering, University of Nebraska--Lincoln, Lincoln, NE 68588, USA}
\author{Md Khirul Alam Titu}
\affiliation[cheme]
{Department of Chemical and Biomolecular Engineering, University of Nebraska--Lincoln, Lincoln, NE 68588, USA}
\author{Srivathsan Badrinarayanan}
\affiliation
{Department of Chemical Engineering, Carnegie Mellon University, 5000 Forbes Avenue, Pittsburgh, PA 15213, USA}
\author{Ravi F. Saraf}
\author{Janghoon Ock}
\email{jock2@unl.edu}
\affiliation[cheme]
{Department of Chemical and Biomolecular Engineering, University of Nebraska--Lincoln, Lincoln, NE 68588, USA}

\begin{document}

\begin{abstract}
Gold nanonecklace networks exhibit non-Ohmic current--voltage (\(I\)--\(V\)) characteristics with a sharp conduction threshold, making them promising for sensing and transistor-like devices. However, their conduction has been understood largely empirically, because experiments cannot see how conduction evolves inside the network. To address this, we develop a graph-based Kirchhoff modeling framework that resolves these internal processes, tracking the potential and current of every junction. Although every active element conducts linearly, the model reproduces the non-Ohmic response $I \propto (V-V_T)^\zeta$. It also captures the experimentally observed effects of junction properties, network density, and voids, which lends confidence to its predicted internal states. These states show that the non-Ohmic response emerges from voltage-driven percolation. Conduction is governed by how junctions are connected rather than by how many are active, with the current scaling approximately as \(N^{2.3}\) while the number of active junctions grows only linearly. Charge-contrast scanning electron microscopy of deposited networks supports this connectivity picture, showing that only a portion of the chains near an electrode are electrically connected to it, while the rest form isolated clusters separated by nanometer-scale gaps whose varying sizes suggest a physical origin for junction activation. Contrary to the expectation that junction heterogeneity strengthens the nonlinear response, homogeneous junctions in a well-connected network produce the most abrupt and reproducible turn-on. Thus, each structural parameter acts as a distinct control knob. These provide design rules for tailoring nanonecklace networks to a desired conduction behavior through synthesis and fabrication.
\end{abstract}


\section{Introduction}
Gold nanonecklace networks are an emerging class of functional electronic nanomaterials characterized by strongly non-Ohmic electric response and a sharp conduction threshold at room temperature~\cite{Prasad2021, Prasad2024}. This threshold voltage, \(V_T\), can be tuned from approximately \(0~\mathrm{V}\) to more than \(5~\mathrm{V}\). This tunability has been exploited in transistors that operate in both air and aqueous environments~\cite{Yu2013}. In addition, the metallic character of the networks provides an effective interface with the biochemical activity of microorganisms, making these materials suitable for biogating applications~\cite{Kane2010}.

These functional properties arise from a hierarchical structure controlled at both the chain and network scales. Gold nanoparticles first self-assemble in solution into quasi-one-dimensional chains, or necklaces~\cite{Saraf2002}, in which ligand chemistry and ion-mediated interactions set the interparticle gaps with subnanometer precision~\cite{ZabetKhosousi2008, Santhanam2016, Lim2025}. These necklaces are then deposited onto a substrate to form two-dimensional monolayer nanoparticle necklace networks (\(N^{3}\)). Electronically, \(N^{3}\) is a multi-tunnel junction (MTJ) structure in which the intrachain tunnel junctions are uniform, whereas the interchain tunneling resistances are polydisperse and dictated by the deposition conditions~\cite{Beloborodov2007, Saraf2010}. The structure therefore offers two complementary control knobs. Solution chemistry tunes the intrachain junctions, while deposition conditions set the interchain junctions together with the connectivity and topology of the network.

At the network scale, the macroscopic response emerges collectively from many junctions. It depends both on their individual junction characteristics and how they are connected into conducting pathways, rather than on the behavior of any single junction. Because these networks are sparse, heterogeneous, and geometrically irregular, macroscopic conduction requires a complete conducting path across the system~\cite{Parthasarathy2004}. Below the threshold voltage, the current remains suppressed. As the applied voltage increases, more junctions become conductive and a percolating pathway emerges at a threshold voltage, \(V_T\), when enough of them connect the electrodes. Above this threshold, the current rises steeply and can be described by the power law, as follows:

\begin{equation}
    I \propto (V - V_T)^{\zeta},
\end{equation}
where \(\zeta\) characterizes the collective growth of conduction above the macroscopic threshold~\cite{Middleton1993}. This growth reflects the continued evolution of the conducting network, as additional junctions create new branches and parallel pathways and redistribute current away from the initial bottlenecks.

Previous experimental and theoretical studies have established several signatures of this collective conduction behavior, including percolation-like conduction onset, power-law current scaling above \(V_T\), and strong sensitivity to network coverage and topology~\cite{Wilson2019, Blunt2007, Parthasarathy2001, Reichhardt2003}. At the junction level, conduction proceeds by electron tunneling across the interparticle gaps. At low temperature, this tunneling is limited by Coulomb blockade, whereas at room temperature gold nanonecklace networks have been shown to cross over to classical percolation, in which the macroscopic response is set by how conducting junctions connect across the network~\cite{Kane2010, Prasad2024}. Consistent with this picture, electron microscopy combined with electric response measurements indicates that the macroscopic electrical response is governed primarily by network topology rather than material composition~\cite{Wilson2019}, as also observed in related branched assemblies~\cite{Tadic2015}.

Despite these advances, the internal dynamics of the percolation remain difficult to access experimentally, especially during device operation. Imaging techniques provide spatially or temporally limited representations of the network state, while conventional two-terminal \(I\)--\(V\) measurements collapse the collective response of the entire network into a single macroscopic curve. Consequently, conventional experimental techniques cannot fully identify which junctions are activated first, track their activation sequence, map the precise topology of the percolation pathways at the conduction threshold, or capture how current is redistributed after individual switching events. Decoupling these junction-level events is essential for understanding how internal networks evolve within \(N^{3}\) systems, a crucial step towards the rational design of advanced nanonecklace architectures for electronic applications. We further image deposited networks by charge-contrast scanning electron microscopy (SEM) at a grounded electrode. The resulting images reveal a hierarchy of connectivity within the network, in which some clusters are electrically connected to the electrode and can transport charge, while others remain isolated and do not contribute to conduction.

To address this gap, we introduce a Kirchhoff-law-based electron-transport model for self-assembled nanonecklace networks. The physical network is represented as a synthetic graph in which nodes and edges describe junctions and their nanonecklace connections, respectively. For a given applied voltage, the model identifies the resulting conducting subgraph and solves Kirchhoff’s circuit laws, treating each node and edge as a resistor, to describe conduction inside the network. In addition to the total current, the model therefore resolves internal quantities, such as node potentials, edge-current magnitudes, and conducting pathways, which are inaccessible to two-terminal measurements. By tracking how the conducting subgraph evolves with voltage, the model links each feature of the strongly non-Ohmic conduction observed in gold nanonecklace networks, namely the threshold, the power-law rise, and the high-bias saturation, to a specific change in network connectivity.

\section{Methods}
\subsection{Nanonetwork Representation}
\begin{figure*}[!hbt]
    \centering
    \includegraphics[width=0.80\textwidth]{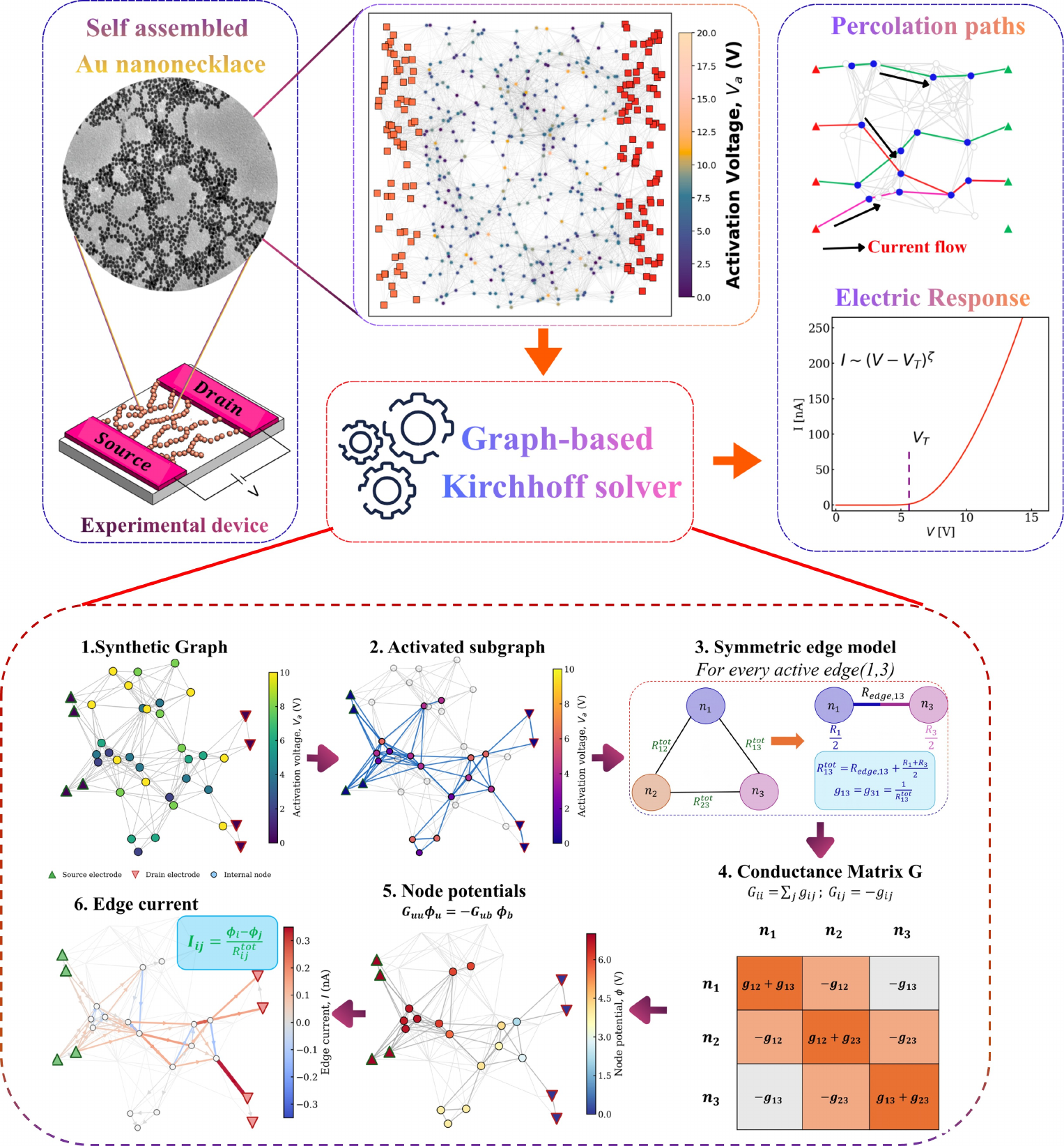}
    \caption{Graph-based modeling framework for gold nanonecklace networks \(N^{3}\). The \(N^{3}\) is represented as a graph whose nodes activate with applied voltage. Kirchhoff's circuit laws are then solved on the resulting conducting subgraph to obtain node potentials, edge currents, conducting pathways, and the \(I\)--\(V\) response.}
    \label{fig:overview}
\end{figure*}

The gold nanonecklace network forms through the self-assembly of individual nanonecklaces, each a quasi-one-dimensional chain\cite{Kane2010, Modak2019}. As shown in the upper-left Transmission Electron Microscopy (TEM) image in Figure~\ref{fig:overview}, the necklaces interconnect at junctions where multiple necklace segments meet, forming a disordered two-dimensional network. Individual necklaces have comparatively uniform nanoparticle spacing, whereas the deposited network exhibits heterogeneous junction spacing.

We represent the system as an undirected geometric graph. Graph nodes denote the junctions at which the strands of the necklace meet, and the graph edges denote the strands of the necklace that connect neighboring junctions. $N$ nodes are placed independently and uniformly within a normalized square domain of side length $L=1$. The nodes $i$ and $j$ are joined when their Euclidean separation satisfies $d_{ij}<r_c$, where the baseline connection radius is $r_c=0.15L$. Nodes with degree zero or one are removed iteratively because they cannot represent internal multisegment junctions. The final number of graph nodes can be slightly smaller than the initial requested value $N$. The source and drain contacts are geometrically assigned after graph construction. The nodes satisfying $x_i\leq0.15L$ are designated source-contact nodes, and the nodes satisfying $x_i\geq0.85L$ are designated drain-contact nodes. The applied voltage \(V\) is imposed on all source-contact nodes, and all drain-contact nodes are grounded at \(0\)~V. Additional graph-generation and sampling details are provided in Supporting Information, Section~S1.

\subsection{Node Activation}

Physically, a junction begins to conduct when electrons can tunnel across its interchain junction gap. In a deposited nanonetwork, this barrier varies from junction to junction with particle size, local spacing, coordination, and dielectric environment~\cite{tran2008charge, tadic2011modeling}. Computing these quantities microscopically for every junction would require detailed knowledge of the local geometry and electrostatics, which is neither experimentally accessible nor computationally tractable for large networks. Instead, the model assigns each junction node $i$ a phenomenological activation voltage $V_{\mathrm{a}, i}$. This captures the junction-to-junction heterogeneity without resolving its microscopic origin, so that the analysis can focus on how the junctions connect into conducting paths and produce the macroscopic \(I\)--\(V\) response.

At applied bias $V$, the activation state of node $i$, \(a_i(V)\), is
\begin{equation}
  a_i(V)=
  \begin{cases}
    1, & V_{\mathrm{a},i}\leq V,\\
    0, & V_{\mathrm{a},i}>V.
  \end{cases}
  \label{eq:activation_rule}
\end{equation}
An edge is electrically available only when both endpoint nodes are active. The activation criterion is determined by the externally applied bias rather than the local node potential, so the model does not compute local fields, capacitances, tunneling probabilities, or charging events.

To represent junction heterogeneity arising from variations in spacing and local environment, activation voltages are sampled independently from a truncated normal distribution truncated to $[0, 20]$~V,
\begin{equation}
  V_{\mathrm{a},i}\sim
  \mathcal{N}\!\left(\langle V_{\mathrm{a}}\rangle,\sigma_a^2\right),
  \qquad 0\leq V_{\mathrm{a},i}\leq20~\mathrm{V}.
  \label{eq:Va_distribution}
\end{equation}
where $\langle V_{\mathrm{a}}\rangle$ and $\sigma_{\mathrm{a}}$ are the mean and standard deviation of the parent distribution (untruncated). Any value falling outside the range \(0\)--\(20\)~V is discarded and redrawn, rather than set to the nearest bound, so that no artificial excess of nodes appears at exactly \(0\) or \(20\)~V. Unless otherwise specified, the voltage is swept from $0$ to $16$~V in increments of $0.5$~V. The parameter values used for each production figure are stated in its caption. The baseline calculation in Figure~\ref{fig:collective_transport} uses a network of $N=500$ nodes with void fraction $f_v=0$, and activation voltages drawn with $\langle V_{\mathrm{a}}\rangle=6~\mathrm{V}$ and $\sigma_{\mathrm{a}}=3~\mathrm{V}$.

\subsection{Resistance Modeling}

Once the conducting subgraph is identified, it is treated as an Ohmic resistor network. The model contains no explicit tunneling-current term and does not evaluate an exponential distance-dependent tunneling law. Instead, each necklace, representing a necklace segment, is assigned an effective resistance proportional to its length,
\begin{equation}
  R_{\mathrm{edge},ij}=k_e d_{ij},
  \label{eq:R_edge}
\end{equation}

where \(k_e=2.0\times10^{10}~\Omega/L\) and \(d_{ij}\) is the length of edge \(ij\) in units of \(L\). This linear scaling reflects the series of uniform intrachain junctions along each necklace. 

Each internal junction  (i.e., activated subgraph node) is also assigned the same fixed resistance,
\begin{equation}
  R_{\mathrm{node}}=3.5\times10^9~\Omega.
  \label{eq:R_node_fixed}
\end{equation}
This resistance is independent of $V_{\mathrm{a},i}$, which prevents changes in the activation-voltage distribution from simultaneously altering both the activation sequence and the post-activation resistance. Thus, $V_{\mathrm{a},i}$ controls when a node becomes available, while $R_{\mathrm{node}}$ controls the resistive penalty for traversing an activated internal junction.

To include node resistances in the circuit, half of the resistance of each endpoint node is assigned to every edge connecting it, so that the total resistance of edge $ij$ is
\begin{equation}
  R_{ij}^{\mathrm{tot}}
  =R_{\mathrm{edge},ij}
  +\frac{1}{2}\left(\widetilde{R}_i+\widetilde{R}_j\right),
  \label{eq:R_total}
\end{equation}
where \( R_{ij}^{\mathrm{tot}} \) is the total resistance between the two nodes, $\widetilde{R}_i=R_{\mathrm{node}}$ for an internal node and $\widetilde{R}_i=0$ for a source or drain contact node. Along an unbranched path, the two half-contributions associated with each internal junction sum to one full junction resistance. This construction retains a symmetric, passive circuit Laplacian and provides a single, unambiguous current for every undirected graph edge (Supporting Information, Section~S2).

\subsection{Graph-Based Kirchhoff Solver}

At each applied voltage \(V\), the nodes satisfying the activation condition in Equation~\eqref{eq:activation_rule}, together with the edges connecting them, form the active subgraph. Only connected components of this subgraph that contain both source and drain contact nodes can carry current across the device. If no such component exists, the device current is set to zero.

Each edge of the conducting subgraph has conductance
\begin{equation}
  g_{ij}=\frac{1}{R_{ij}^{\mathrm{tot}}},
  \label{eq:conductance}
\end{equation}
where \(R_{ij}^{\mathrm{tot}}\) is given by Equation~\eqref{eq:R_total}. These conductances are assembled into the conductance matrix (weighted graph Laplacian) \(\mathbf{G}\), with entries
\begin{equation}
  G_{ii}=\sum_{j\in\mathcal{N}(i)}g_{ij},
  \qquad
  G_{ij}=-g_{ij}\quad(i\neq j),
  \label{eq:laplacian_entries}
\end{equation}
where \(\mathcal{N}(i)\) is the set of active neighbors of node \(i\). Because each edge has the same conductance in either direction, \(g_{ij}=g_{ji}\), the matrix \(\mathbf{G}\) is symmetric.

All source contact nodes are held at the applied voltage, and all drain contact nodes are grounded,

\begin{equation}
  \phi_s=V \;\; \text{for all } s\in\mathcal{S},
  \qquad
  \phi_d=0 \;\; \text{for all } d\in\mathcal{D},
  \label{eq:boundary_conditions}
\end{equation}
where \(\mathcal{S}\) and \(\mathcal{D}\) are the sets of source and drain contact nodes. Partitioning the nodes into internal nodes with unknown potentials (\(u\)) and contact nodes with fixed potentials (\(b\)), Kirchhoff's current law at each internal node gives the linear system
\begin{equation}
  \mathbf{G}_{uu}\boldsymbol{\phi}_u
  =
  -\mathbf{G}_{ub}\boldsymbol{\phi}_b,
  \label{eq:linear_system}
\end{equation}
where \(\boldsymbol{\phi}_u\) and \(\boldsymbol{\phi}_b\) are the vectors of internal and contact node potentials, and \(\mathbf{G}_{uu}\) and \(\mathbf{G}_{ub}\) are the corresponding blocks of \(\mathbf{G}\).

Once the node potentials are known, the current on each edge follows from Ohm's law,
\begin{equation}
  I_{ij}
  =
  g_{ij}(\phi_i-\phi_j)
  =
  \frac{\phi_i-\phi_j}{R_{ij}^{\mathrm{tot}}},
  \label{eq:edge_current}
\end{equation}
where a positive \(I_{ij}\) denotes current flowing from node \(i\) to node \(j\). The total device current is evaluated independently at both electrodes as the net current leaving the source nodes, \(I_{\mathrm{s}}=\sum_{s\in\mathcal{S}}\sum_{j\in\mathcal{N}(s)} I_{sj}\), and the net current entering the drain nodes, \(I_{\mathrm{d}}=\sum_{d\in\mathcal{D}}\sum_{j\in\mathcal{N}(d)} I_{jd}\). The reported current is their average,
\begin{equation}
  I(V)
  =
  \frac{|I_{\mathrm{s}}|+|I_{\mathrm{d}}|}{2}.
  \label{eq:device_current}
\end{equation}

Because current is conserved, \(I_{\mathrm{s}}\) and \(I_{\mathrm{d}}\) are equal in magnitude, and their difference serves as a numerical check (Supporting Information, Section~S3). The effective device conductance is

\begin{equation}
  G_{\mathrm{eff}}(V)=\frac{I(V)}{V},
  \label{eq:effective_conductance}
\end{equation}
and is defined as zero at \(V=0\).
 
The model does not assign currents to independent pathways or sum currents over enumerated paths. Additional implementation details, numerical tolerances, and a worked three-node calculation are provided in Supporting Information, Section~S3.

\subsection{Electrical Response Fitting}
The threshold voltage, $V_T$, is defined as the first sampled voltage at which a conducting subgraph spans the source and drain,
\begin{equation}
  V_T\equiv V_{\mathrm{perc}}
  =\min\{V_k\in\mathcal{V} : I(V_k)>0\},
  \label{eq:VT_definition}
\end{equation}
where \(\mathcal{V}=\{0, 0.5, \ldots, 16\}\)~V is the set of sweep voltages. Since all resistances are positive, \(I(V_k)>0\) holds exactly when \(V_k>0\) and a conducting subgraph exists.

\begin{equation}
  I = I_0\, (V - V_T)^{\zeta},  \qquad V > V_T,
  \label{eq:power_law}
\end{equation}
where \(I_0\) is a prefactor and \(\zeta\) is the scaling exponent. We obtain $\zeta$ by linear regression of \(\log I\) against \(\log(V-V_T)\). The threshold point itself is excluded because \(V-V_T=0\) there, and the fitting window ends before the first voltage at which $90\%$ of the nodes are active. The fitting procedure is detailed in Supporting Information, Section~S4.

\subsection{Void Placement Strategy}

Depending on the deposition conditions, the fabrication process can leave particle-depleted regions in real nanonecklace networks, interrupting otherwise direct conducting routes~\cite{Parthasarathy2001}. These regions are represented as circular exclusion zones that contain no graph nodes and cannot be crossed by graph edges. Each void has radius $r_v=0.08L$. For a prescribed void fraction $f_v$, the number of circles is
\begin{equation}
  n_v=\operatorname{round}\!\left(\frac{L^2f_v}{\pi r_v^2}\right).
  \label{eq:void_count}
\end{equation}
Void overlap is allowed, and the achieved union-area fraction is calculated separately so that overlapping regions are counted only once. The placement, edge-removal, and achieved-area calculations are described in Supporting Information, Section~S7.

The node coordinates are generated by rejection sampling so that all accepted nodes lie outside the voids. Edges are first proposed using the applicable connection radius and are removed if their straight-line segments intersect any void. After edge removal, degree-zero and degree-one nodes are pruned iteratively. 

\subsection{Parameter Sweep Design}

Systematic parameter sweeps are designed to isolate the effect of four network parameters on characteristic $I$--$V$ (Table~\ref{tab:parameter_sweeps}). Each parameter represents distinct physical features of a real nanonecklace network. The mean of the activation voltage distribution, \(\langle V_{\mathrm{a}}\rangle\), reflects the typical interparticle spacing and local environment of the junctions, and thus the typical bias required for a junction to conduct. The standard deviation, \(\sigma_a\), reflects the junction-to-junction heterogeneity in these properties introduced during self-assembly and fabrication. The node count, \(N\), represents the amount or coverage of the nanonecklace material deposited between the electrodes, and the void fraction, \(f_v\), represents particle-free regions in the deposited network that can be caused during the fabrication process. To characterize the variability arising from the random network realization, each parameter combination was evaluated with the same set of five random seeds.

\begin{table}[htb]
\centering
\caption{Parameter sweep ranges. Each row varies one parameter while the others are held at the listed fixed values.}
\label{tab:parameter_sweeps}
\resizebox{\linewidth}{!}{%
\begin{tabular}{@{}lllll@{}}
\toprule
\textbf{Sweep target} & \textbf{$\langle V_{\mathrm{a}} \rangle$ [V]} & \textbf{$N$ [--]} & \textbf{$f_v$ [--]} & \textbf{$\sigma_a$ [V]} \\ \midrule
Mean of activation voltage $\langle V_{\mathrm{a}} \rangle$ & 4, 6, 8, 10 & 500 (fixed) & 0.00 (fixed) & 1 (fixed) \\
Number of nodes $N$ & 6 (fixed) & 200, 400, 600, 800 & 0.00 (fixed) & 3 (fixed) \\
Void fraction $f_v$ & 6 (fixed) & 500 (fixed) & 0.00, 0.05, 0.10, 0.15, 0.20, 0.25 & 3 (fixed) \\
Std. of activation voltage $\sigma_a$ & 8 (fixed) & 500 (fixed) & 0.00 (fixed) & 1, 3, 5, 7 \\ \bottomrule
\end{tabular}
}
\end{table}

\subsection{Charge-Contrast SEM}

Au nanoparticle chains were assembled in solution by ion-mediated directed self-assembly, with Zn(NO\(_3\))\(_2\) as the divalent salt~\cite{Lim2025}, and deposited onto the substrate following Wilson \textit{et al.}~\cite{Wilson2019}. The substrate was a Si chip with thermal oxide and a 100~nm thick Au electrode patterned by photolithography. The chip was cleaned in an oxygen plasma (80~W, 100~mTorr, 2~min), spin-coated with poly(allylamine hydrochloride) from a 1~mg~mL\(^{-1}\) aqueous solution, and rinsed with deionized water. The chain suspension (427~\(\mu\)L) was diluted with 713~\(\mu\)L of deionized water and deposited by centrifugation at approximately 1900~\(g\) for 10~min, after which the chip was rinsed with water and air-dried.

SEM images were acquired on an FEI Helios NanoLab 660 using the through-the-lens detector in secondary-electron mode (5~kV, 0.1~nA, 3~\(\mu\)s dwell time, 3.1~mm working distance), as single \(3072\times2048\)-pixel frames at 1.04~nm per pixel. The Au electrode was grounded, and detector settings were held fixed within each frame.

Image analysis was performed on the full original frames in Python with scikit-image~\cite{vanderWalt2014}. Brightness is the recorded secondary-electron signal, scaled from the 16-bit detector range to 0--1. Noise was reduced by non-local-means denoising~\cite{Buades2005} with filter strength \(h=0.8\sigma\), where \(\sigma\) is the estimated noise standard deviation~\cite{Donoho1994}. The images displayed in Figure~\ref{fig:sem}a,e were more strongly denoised (\(h=1.4\sigma\)) and are shown on a linear gray scale without further adjustment. The slowly varying substrate background was estimated by a morphological opening with a 12-pixel disk, which removes features narrower than approximately 25~nm, followed by Gaussian smoothing, and was subtracted from the image (top-hat transform; Figure~\ref{fig:sem}c). Pixels more than 0.07 above this background, or saturated, were classified as chain, and specks smaller than 60 pixels and holes smaller than 40 pixels were removed (Figure~\ref{fig:sem}d).

The foot of the electrode sidewall was traced in each image row as the position of the steepest rise in substrate brightness, and the sloped sidewall was excluded from brightness comparisons. Chain pixels were grouped into 8-connected clusters, and clusters reaching the electrode were classified as connected and all others as isolated (Figure~\ref{fig:sem}b)~\cite{Wilson2019}. The mean brightness of each isolated cluster of at least 150 pixels on the substrate was compared with the median brightness of connected chain pixels at the same distance from the electrode (\(\pm0.1~\mu\)m).

The gap of a cluster was defined as the shortest distance across bare substrate between its pixels and the nearest pixel of another cluster, and gaps of 10~nm or less were treated as possible tunnel junctions. Routes in Figure~\ref{fig:sem}e were found by a shortest-path search from the electrode (Dijkstra's algorithm)~\cite{Dijkstra1959}, with a low cost for moving along chains, a high cost for crossing a gap, and bare substrate excluded, so that each route crosses as few gaps as possible. The number of junctions on each route was unchanged when the gap cost was increased 25,000-fold. A second image, recorded at a different location along the same electrode (0.2~nA, 10~\(\mu\)s dwell time, 1.35~nm per pixel), was analyzed in the same way, with pixel-based sizes rescaled to the same physical dimensions. Further details are given in Supporting Information, Section~S9.

\section{Results and Discussion}
\subsection{Emergent Non-Ohmic Conduction}

\begin{figure*}[!hbt]
\centering
\includegraphics[width=.99\linewidth]{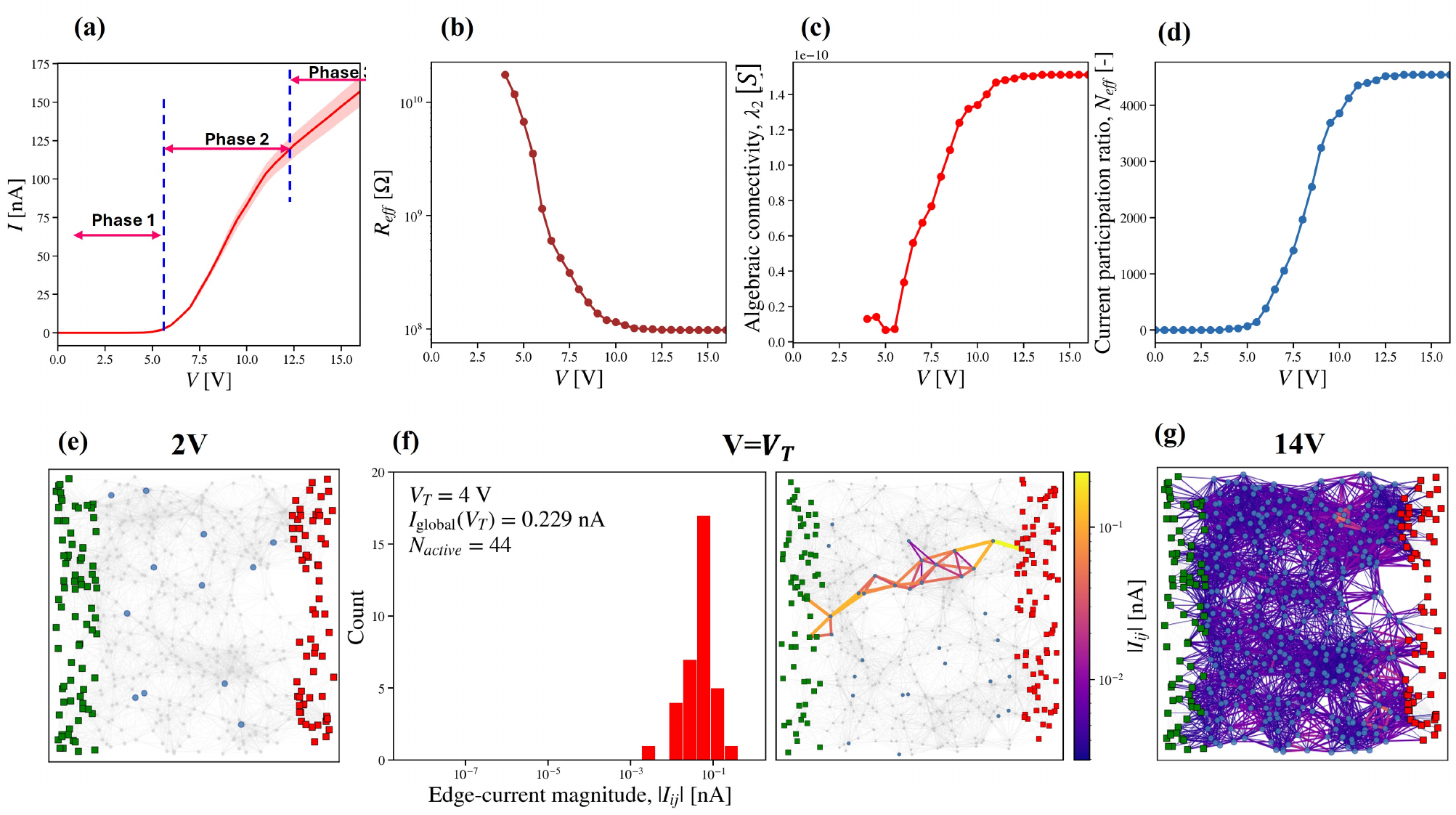}
\caption{Macroscopic electrical response and internal subnetwork evolution. (a)~Ensemble-averaged \(I\)--\(V\) response for five network realizations at \(N=500\), \(\langle V_{\mathrm{a}}\rangle=6\)~V, \(\sigma_{\mathrm{a}}=3\)~V, and \(f_v=0.00\). The shaded band shows the variation. (b)~Effective source--drain resistance \(R_{\mathrm{eff}}\). (c)~Conductance-weighted algebraic connectivity \(\lambda_2\). (d)~Current participation ratio \(N_{\mathrm{eff}}\), a measure of how many conducting edges effectively share the electrical current. (e)~Active network at \(2\)~V, below \(V_T\). (f)~Edge-current distribution and current-carrying network  at \(V_T=V_{\mathrm{perc}}=4\)~V. The histogram and network show the same set of edges. (g)~Conducting network of the same realization at \(14\)~V. Source (green) and drain (red) nodes are marked; edge color indicates current magnitude (log scale).}
\label{fig:collective_transport}
\end{figure*}

As bias is applied, the network passes through three electron transport phases (Figure~\ref{fig:collective_transport}a): a low-bias blockade (Phase~1), a nonlinear growth phase that begins at the threshold voltage (Phase~2), and a nearly saturated, quasi-Ohmic phase (Phase~3). Similar electric behavior has been predicted for disordered arrays of metallic particles~\cite{Middleton1993, Bascones2008, Narumi2011}. 

In Phase~1, the applied voltage activates only a small fraction of the junction nodes. At \(2\)~V, several nodes are active, but form disconnected groups that do not join the source to the drain, so no current crosses the window (Figure~\ref{fig:collective_transport}e). Activating individual junctions is not sufficient for conduction; the active junctions must also form a continuous source--drain path. The low-bias blockade is consequently a network-connectivity effect, consistent with percolation-controlled electron transport in disordered nanoparticle arrays~\cite{Parthasarathy2001, Reichhardt2003, Wilson2019}.

For the exemplary visualization, the first conducting connection appears at \(V_T=4\)~V (Figure~\ref{fig:collective_transport}f). Only \(44\) of the \(500\) nodes are active at this voltage, yet they form a narrow backbone that carries \(0.229\)~nA, so conduction begins. Current along this backbone is also highly uneven. The edge-current histogram and the corresponding snapshot show that a few edges carry large currents and form the main backbone, whereas many others carry much smaller currents along secondary branches. Therefore, electron transport is strongly localized at the onset of percolation, consistent with current localization and bias-dependent changes observed experimentally in nanoparticle networks~\cite{Blunt2007, Ong2013, Wilson2019}.

The nonlinear current growth in Phase~2 can be understood by writing the device current as:
\begin{equation}
  I(V)=\frac{V}{R_{\mathrm{eff}}(V)},
  \label{eq:I_Reff}
\end{equation}
where \(R_{\mathrm{eff}}=1/G_{\mathrm{eff}}\) is the effective resistance of the network (Equation~\eqref{eq:effective_conductance}). If the conducting subgraph does not change, \(R_{\mathrm{eff}}\) would be constant and the current would grow linearly. Instead, increasing \(V\) activates additional nodes and edges that attach to the initial backbone, and network metrics track this evolution. \(R_{\mathrm{eff}}\) decreases by more than two orders of magnitude before leveling off (Figure~\ref{fig:collective_transport}b), as new connections open easier, parallel routes between the electrodes. In addition, algebraic connectivity \(\lambda_2\) increases and eventually plateaus (Figure~\ref{fig:collective_transport}c; Supporting Information, Section~S5), indicating that the initially fragile backbone becomes more strongly connected. 

In Phase~3, the conducting structure barely changes. The plateau in \(N_{\mathrm{eff}}\) shows that the current redistribution among the conducting edges is nearly saturated (Figure~\ref{fig:collective_transport}d; Supporting Information, Section~S6). The \(14\)~V snapshot shows a dense network in which activation is nearly saturated (Figure~\ref{fig:collective_transport}g). The additional voltage now increases the current mainly through existing connections, so \(R_{\mathrm{eff}}\) in Equation~\eqref{eq:I_Reff} is approximately constant and the response becomes quasi-Ohmic. Because this region is no longer dominated by the activation of new conducting routes, it is excluded from the power-law fit.

All these simulated results link the \(I\)--\(V\) response to the internal state of the network. The non-Ohmic response of \(N^{3}\) arises because the conducting subgraph changes with voltage through bias-induced percolation and current redistribution, rather than because a larger voltage is applied across a fixed resistor network~\cite{Middleton1993, Parthasarathy2001, Parthasarathy2004, Reichhardt2003, Bascones2008, Wilson2019, Prasad2021, Prasad2024}.

\subsection{Activation-Induced Threshold Shift}

\begin{figure*}[!hbt]
\centering
\includegraphics[width=.9\linewidth]{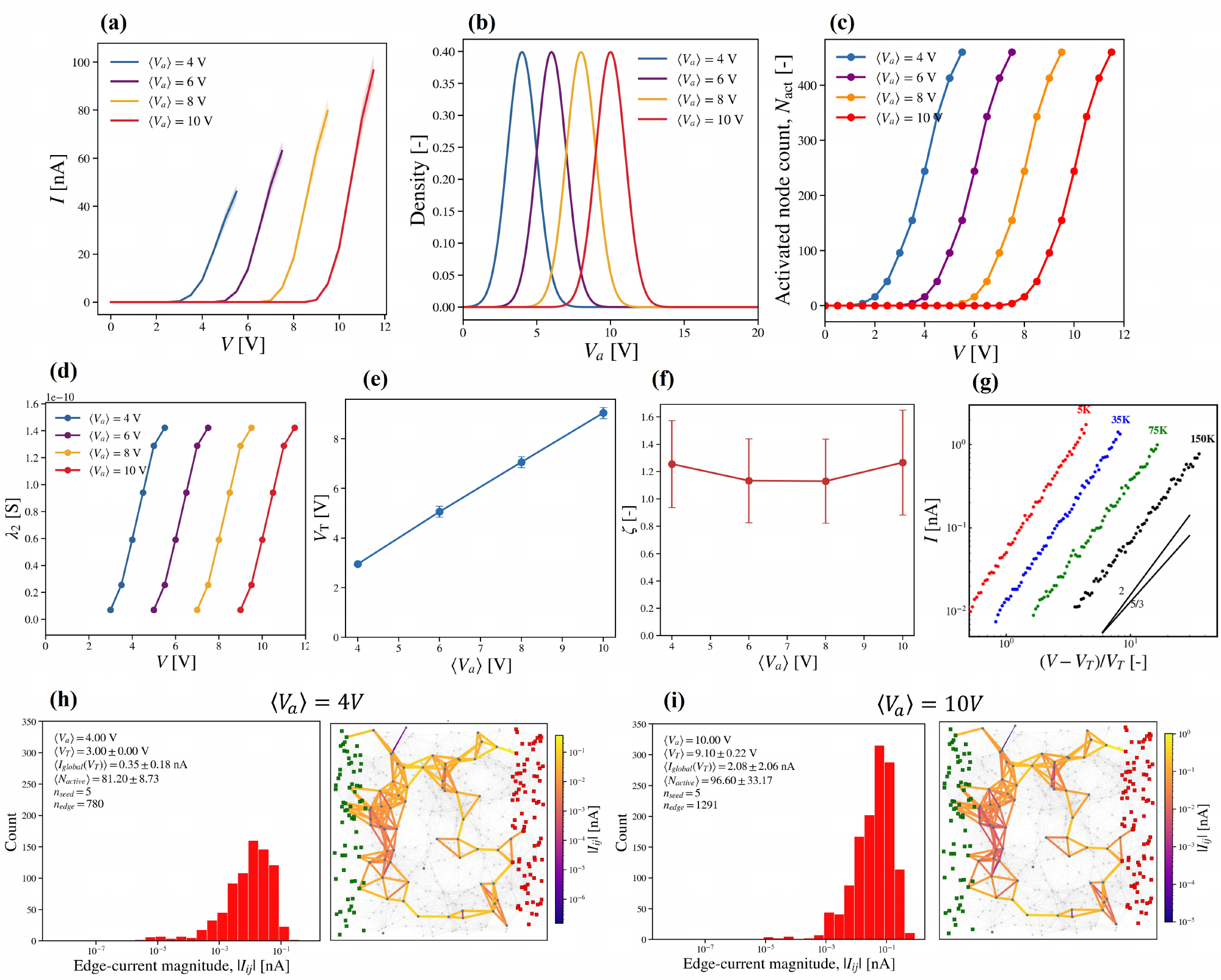}
\caption{Effect of the mean activation voltage on conduction. (a)~\(I\)--\(V\) characteristics for \(\langle V_{\mathrm{a}}\rangle=4\), \(6\), \(8\), and \(10\)~V (\(\sigma_{\mathrm{a}}=1\)~V, \(N=500\), \(f_v=0.00\)), with (b)~the corresponding activation-voltage distributions, (c)~activated-node population \(N_{\mathrm{act}}\) , and  (d)~conductance-weighted algebraic connectivity \(\lambda_2\). (e)~Threshold voltage \(V_T\) and (f)~Scaling exponent \(\zeta\) versus \(\langle V_{\mathrm{a}}\rangle\), shown as means and standard deviations over five realizations. (g)~Temperature-dependent scaled \(I\)--\(V\) characteristics of experimental self-assembled gold nanonecklace networks, reproduced from Kane \textit{et al.}~\cite{Kane2010}. (h, i)~Edge-current distribution and conducting networks at threshold with \(\langle V_{\mathrm{a}}\rangle=4\)~V and  \(\langle V_{\mathrm{a}}\rangle=10\)~V.}
\label{fig:meanVa_shift}
\end{figure*}

The mean activation voltage, \(\langle V_{\mathrm{a}}\rangle\), sets when the network begins to conduct while not changing how it conducts. As \(\langle V_{\mathrm{a}}\rangle\) increases from \(4\) to \(10\)~V, the \(I\)--\(V\) curves shift rigidly to higher voltage while retaining their shape (Figure~\ref{fig:meanVa_shift}a), and \(V_T\) increases nearly proportionally with \(\langle V_{\mathrm{a}}\rangle\), from \(3.00\pm0.00\)~V to \(9.10\pm0.22\)~V (Figure~\ref{fig:meanVa_shift}e). Consistent with the unchanged curve shape, the scaling exponent \(\zeta\) remains approximately constant across the sweep (Figure~\ref{fig:meanVa_shift}f).

This shift originates from the activation process itself. Because each node becomes active once the applied voltage reaches its \(V_{\mathrm{a}, i}\), increasing \(\langle V_{\mathrm{a}}\rangle\) translates the entire activation-voltage distribution to higher voltage (Figure~\ref{fig:meanVa_shift}b), and the activated-node population and connectivity follow by the same increment (Figure~\ref{fig:meanVa_shift}c, d). Percolation is therefore simply delayed, but the network state at which it occurs is essentially unchanged. At threshold, a comparable number of nodes is active for \(\langle V_{\mathrm{a}}\rangle=4\)~V (\(81\pm9\)) and \(10\)~V (\(97\pm33\)), and both conducting networks consist of a sparse backbone spanning the electrodes (Figure~\ref{fig:meanVa_shift}h, i). Since \(V_{\mathrm{a}, i}\) only determines when a node becomes available and does not alter the resistance of active connections, this shift results in delayed conduction. In physical networks, a higher effective activation requirement could arise from larger interparticle separations or changes in the dielectric environment~\cite{ZabetKhosousi2008, RevModPhys.79.469, Qu2017}, suggesting that \(V_T\) can be tuned through the properties of the local junction or medium without altering the collective character of the transition.

This separation between threshold position and response shape agrees with temperature-dependent measurements, which provide a direct experimental analogue of the \(\langle V_{\mathrm{a}}\rangle\) sweep. Because the same network is measured at every temperature, its topology remains fixed, while thermal energy helps electrons overcome the local charging barriers and thus lowers the effective threshold of every junction~\cite{Kane2010}. Varying temperature therefore shifts the junction-threshold scale without changing the network, as varying \(\langle V_{\mathrm{a}}\rangle\) does in the model. Accordingly, the experimental curves in Figure~\ref{fig:meanVa_shift}g shift in onset while their slope on the logarithmic plot, and hence the scaling exponent, remains similar, supporting the interpretation of \(\langle V_{\mathrm{a}}\rangle\) as an effective local activation scale. Additional voltage-dependent activation snapshots and seed-to-seed threshold pathways are shown in Supporting Information, Section~S8.

This agreement indicates that separating junction activation from junction resistance captures the main experimental trends. Because increasing \(\langle V_{\mathrm{a}}\rangle\) raises the turn-on voltage of every junction, some shift of the \(I\)--\(V\) curve is expected by design. The model does not, however, prescribe \(V_T\) or \(\zeta\). The nearly one-to-one dependence of \(V_T\) on \(\langle V_{\mathrm{a}}\rangle\) and the unchanged \(\zeta\) both emerge from how the active junctions connect, and both agree qualitatively with experiments~\cite{Kane2010, Parthasarathy2004}. The junction-level physics can therefore be absorbed into an effective activation voltage, while the macroscopic response is set by when junctions become conductive and how they are connected.

\subsection{Conduction Scaling with Density}

To separate the effect of connectivity from that of the number of active nodes, \(N\) is varied from \(200\) to \(800\) while the simulation area, connection radius, and activation-voltage distribution are held fixed. Increasing \(N\) therefore places more nodes within the same area and raises the network density. 

\begin{figure*}[!hbt]
\centering
\includegraphics[width=.9\linewidth]{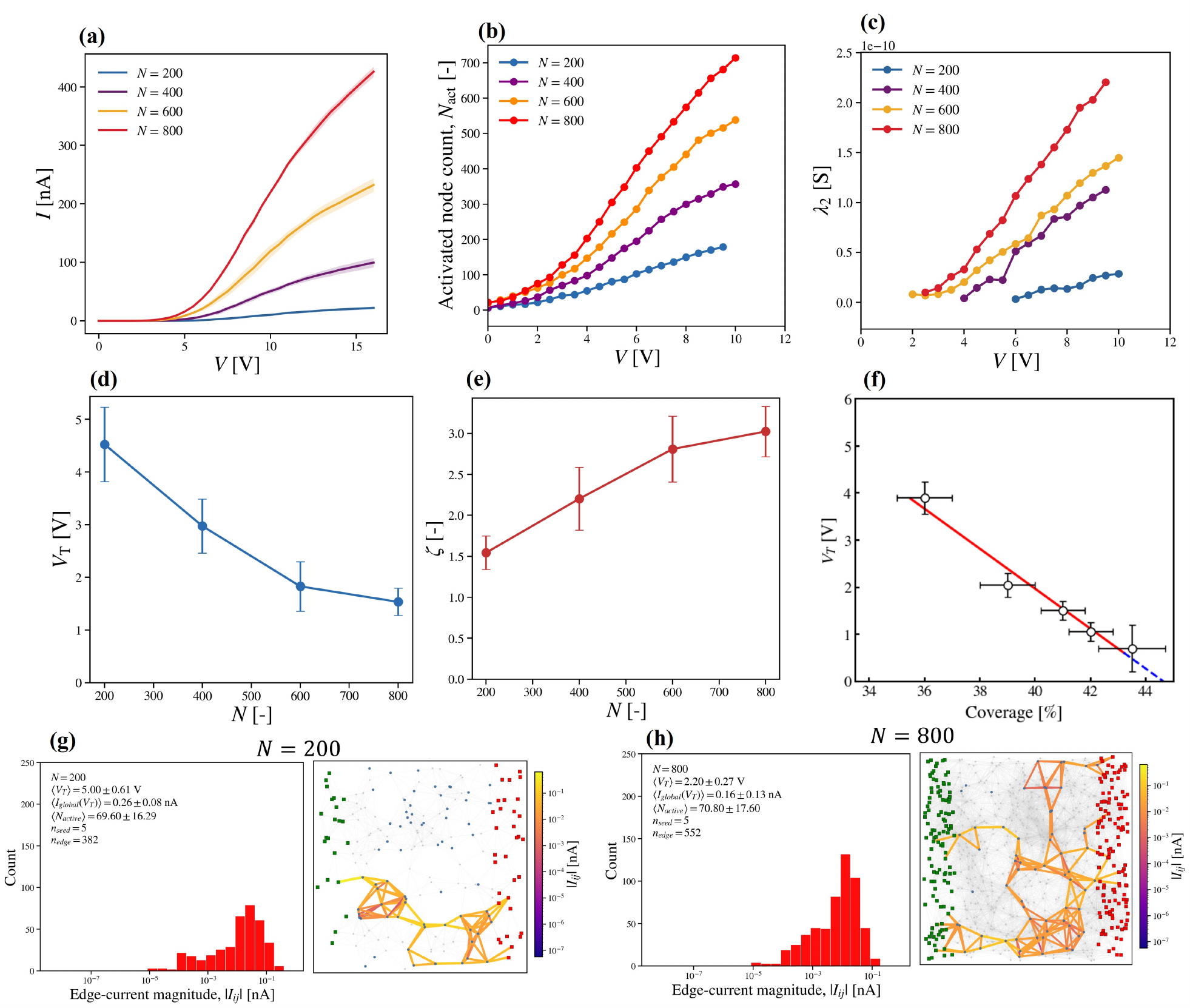}
\caption{Effect of network density on conduction. (a)~\(I\)--\(V\) characteristics for \(N=200\), \(400\), \(600\), and \(800\) (\(\langle V_{\mathrm{a}}\rangle=6\)~V, \(\sigma_a=3\)~V), with the corresponding (b)~activated-node count \(N_{\mathrm{act}}\) and (c)~conductance-weighted algebraic connectivity \(\lambda_2\). (d)~Threshold voltage \(V_T\) and (e)~scaling exponent \(\zeta\) versus \(N\), shown as means and standard deviations over five realizations. (f)~Experimental dependence of threshold voltage on nanonecklace coverage, reproduced from Wilson \textit{et al.}~\cite{Wilson2019}. (g, h)~Edge-current distributions and conducting networks at threshold for representative \(N=200\) and \(N=800\); annotated values are ensemble statistics.}
\label{fig:N_sweep}
\end{figure*}

The current grows much faster than the number of active nodes. At a common applied voltage of \(10\)~V, the current increases by factors of approximately \(5.4\), \(12.7\), and \(23.6\) for \(N=400\), \(600\), and \(800\), respectively, relative to \(N=200\) (Figure~\ref{fig:N_sweep}a), corresponding empirically to \(I\propto N^{2.3}\) over the simulated range. In contrast, \(N_{\mathrm{act}}\) increases approximately in proportion to \(N\). Quadrupling \(N\) yields roughly four times as many active nodes but more than twenty times as much current (Figure~\ref{fig:N_sweep}a, b). 

The difference follows from the geometry of the network. For randomly placed nodes at density \(\rho=N/A\), each node has on average \(\langle k\rangle\approx\rho\pi r_c^2\) neighbors within the connection radius \(r_c\), so the number of edges scales as
\begin{equation}
  E\approx\frac{N\langle k\rangle}{2}\approx\frac{\pi r_c^2}{2A}N^{2}.
  \label{eq:edge_scaling}
\end{equation}
Adding nodes in a fixed area thus increases the number of connections quadratically, providing more parallel routes and reducing the dependence on individual bottlenecks. This near-quadratic growth in connectivity is consistent with the observed \(I\propto N^{2.3}\).

The network metrics confirm this stronger connectivity. As \(N\) increases, \(\lambda_2\) rises more rapidly with voltage and reaches a larger value  (Figure~\ref{fig:N_sweep}c). Because \(\lambda_2\) is calculated from the conductance-weighted Laplacian without normalization, this increase reflects the combined effects of greater node density and mean degree. The representative networks make this structural difference visible (Figure~\ref{fig:N_sweep}g, h). At \(N=200\), current flows through a narrow backbone with few branches, whereas at \(N=800\) it is distributed over several interconnected routes and a much larger population of current-carrying edges. Additional voltage-dependent density snapshots and seed-to-seed threshold pathways are shown in Supplementary Figures~S3 and S4.

Higher density also lowers the voltage required for percolation, with the mean \(V_T\) decreasing from approximately \(4.5\)~V at \(N=200\) to approximately \(1.5\)~V at \(N=800\) (Figure~\ref{fig:N_sweep}d). A sparse network must activate a large fraction of its nodes before a continuous source--drain path forms, whereas in a dense network the small set of earliest-activated nodes is already likely to contain a spanning route (Figure~\ref{fig:N_sweep}g, h). This is consistent with models in which the threshold is set by the first continuous low-barrier path across a disordered system~\cite{Middleton1993, Elteto2004, Parthasarathy2004}. The same trend is observed experimentally. Wilson \textit{et al.} found that increasing nanonecklace coverage lowers the measured threshold voltage~\cite{Wilson2019} (Figure~\ref{fig:N_sweep}f), which the model reproduces with \(N\) serving as a proxy for coverage~\cite{Reichhardt2003, Prasad2024}.

The scaling exponent increases with density, from approximately \(\zeta=1.5\) at \(N=200\) to approximately \(\zeta=3.0\) at \(N=800\) (Figure~\ref{fig:N_sweep}e). This indicates that the current rises more steeply above \(V_T\) in denser networks. Once percolation begins, each newly activated nodes in a dense network can connect through many existing neighbors, so the conducting network expands rapidly; in a sparse network, newly activated nodes have fewer opportunities to join the source--drain backbone. This dependence of the nonlinear scaling on network structure is consistent with previous studies of disordered nanoparticle arrays, which showed that the nonlinear scaling depends on how conduction channels open and connect across the network~\cite{Middleton1993, Parthasarathy2001, Reichhardt2003}.

\subsection{Topological Control of Conduction}

\begin{figure*}[!hbt]
\centering
\includegraphics[width=.9\linewidth]{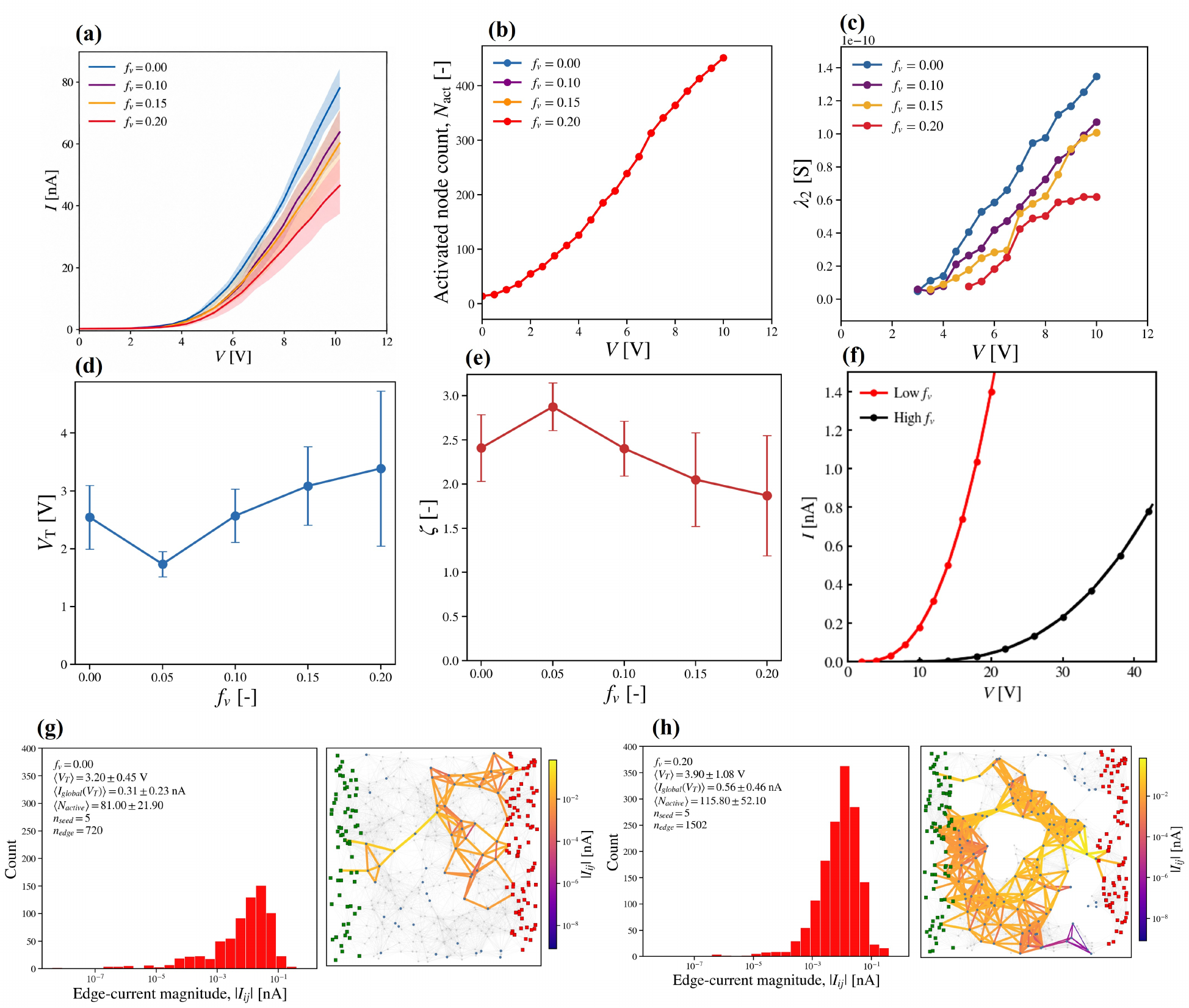}
\caption{Effect of void on conduction. (a)~\(I\)--\(V\) characteristics for \(f_v=0.00\), \(0.10\), \(0.15\), and \(0.20\) (\(N=500\), \(\langle V_{\mathrm{a}}\rangle=6\)~V, \(\sigma_a=3\)~V), with the corresponding (b)~activated-node count \(N_{\mathrm{act}}\) and (c)~conductance-weighted algebraic connectivity \(\lambda_2\). (d)~Threshold voltage \(V_T\) and (e)~scaling exponent \(\zeta\) versus \(f_v\), shown as means and standard deviations over five realizations. (f)~Experimental \(I\)--\(V\) characteristics of gold nanonecklace networks with comparable coverage but different void contents, reproduced from Wilson \textit{et al.}~\cite{Wilson2019}. (g, h)~Edge-current distributions and conducting networks at threshold for representative \(f_v=0.00\) and \(f_v=0.20\) states (annotated values are ensemble statistics).}
\label{fig:void}
\end{figure*}

Depending on the fabrication conditions, networks with the same number of nodes and the same activation-voltage distribution can still differ in topology, and therefore in their electrical response. To isolate this effect, voids are introduced while \(N=500\), \(\langle V_{\mathrm{a}}\rangle=6\)~V, and \(\sigma_a=3\)~V are held fixed. The void fraction \(f_v\) therefore changes, where nodes and edges can occur without changing how many nodes there are or how they activate. Voids reduce the current, most clearly for \(f_v=0.20\), which carries less current than the void-free network throughout most of the nonlinear region (Figure~\ref{fig:void}a).

This reduction arises from connectivity rather than the number of active nodes. The activated-node \(N_{\mathrm{act}}\) curves overlap exactly for all void fractions (Figure~\ref{fig:void}b). Any difference in conduction must come from how the active nodes are arranged and connected. The algebraic connectivity shows this structural difference directly. The void-free network develops the largest \(\lambda_2\), while the void-containing networks generally develop smaller values (Figure~\ref{fig:void}c), indicating weaker connections and more pronounced bottlenecks. This is because voids remove direct routes, divide the available space, and force conducting paths to bend around excluded regions~\cite{Parthasarathy2001}. Accordingly, the void-free network forms relatively direct branches, whereas the \(f_v=0.20\) network bends around excluded regions into a more irregular backbone  (Figure~\ref{fig:void}g,h).

Weakened connectivity also delays percolation. The mean \(V_T\) increases from approximately \(2.6\)~V for the void-free network to approximately \(3.4\)~V at \(f_v=0.20\) (Figure~\ref{fig:void}d), because more nodes must activate before a path can connect the electrodes while avoiding the excluded regions~\cite{Middleton1993, Elteto2004, Parthasarathy2004}. Since more nodes are already active and the bias is higher when this path finally forms, the onset current is larger for \(f_v=0.20\), even though its current at a common bias is lower. 

These simulation results agree qualitatively with experiments. Wilson \textit{et al.} showed that gold nanonecklace films with similar surface coverage can have different threshold voltages and currents because of differences in void content and pathway tortuosity, with the high-void network requiring a larger voltage to conduct~\cite{Wilson2019}. The simulations explain this observation at the network level. Voids reduce algebraic connectivity, introduce bottlenecks, delay the formation of a spanning component, and restrict the subsequent expansion of the conducting network.

The broad error bars in Figure~\ref{fig:void}a, d, and e reflect the sensitivity of conduction to the exact placement of the voids. Beyond the randomness in node positions and activation voltages present in every realization, void-containing networks add a further source of variability: the location and geometry of the voids themselves. Therefore, uniform fabrication is important for reliable device performance. Supplementary Figures~S5 and S6 show additional void-dependent activation snapshots and seed-to-seed conduction pathways.

\subsection{Experimental Evidence for a Hierarchy of Connectivity}

The simulations make two predictions that can be examined in deposited networks. First, conduction is governed by how junctions are connected rather than by how much material is present, so chains near an electrode should contribute only if they are joined to it (Figures~\ref{fig:N_sweep} and \ref{fig:void}). Second, the conducting network grows stepwise as junctions with different activation voltages join it one after another (Figure~\ref{fig:collective_transport}). Charge-contrast SEM at a grounded electrode allows both to be tested~\cite{Wilson2019}. The portion of the network connected to the electrode drains the charge injected by the electron beam and appears bright, whereas isolated clusters accumulate charge and appear darker~\cite{Cazaux2008}, so image brightness maps which clusters can join the conduction path.

\begin{figure*}[!hbtp]
\centering
\includegraphics[width=.8\linewidth]{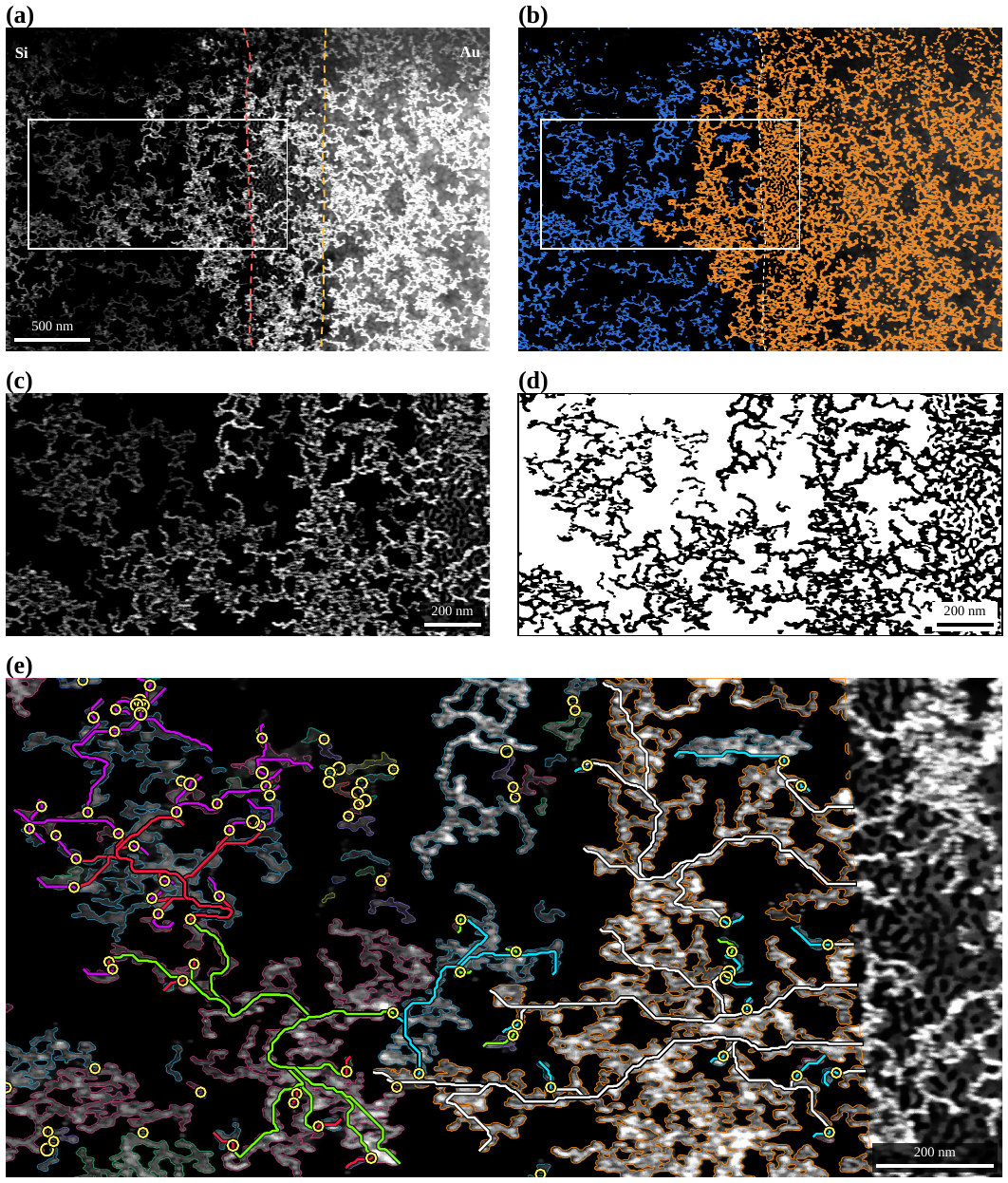}
\caption{Charge-contrast SEM image of a necklace network at a grounded electrode. (a)~Necklace network at the edge of a grounded Au electrode (5~kV, through-the-lens secondary-electron detector, 1.04~nm per pixel). Dashed lines mark the foot (red) and top (yellow) of the electrode sidewall, and the box marks the area shown in (c)--(e). (b)~Chains connected to the electrode (orange) and isolated clusters (blue). (c)~Background-subtracted (top-hat) image of the boxed area and (d)~the binary chain mask obtained from it (chains black), on which all connectivity measurements are based. (e)~Hierarchy of connectivity (electrode at right). The network connected to the electrode is outlined in orange, and each isolated cluster is shown in its own color. Yellow circles mark nanometer-scale gaps between clusters, treated as possible tunnel junctions. Lines trace the route from the electrode to each reachable cluster that crosses the fewest such gaps, colored white within the connected network and cyan, green, red, and magenta after one, two, three, and four or more junctions, respectively. Scale bars, 500~nm (a) and 200~nm (c--e).}
\label{fig:sem}
\end{figure*}

The SEM images confirm that much of the deposited material is not connected to the electrode. Only about one-third of the chain area on the substrate is joined to the electrode, and the remaining chains form darker, isolated clusters (Figure~\ref{fig:sem}a,b). Although many of these clusters lie close to the connected network, their gaps are wide enough that the injected charge cannot tunnel to it. At similar distances from the electrode, isolated clusters are darker than connected chains in 148 of 154 cases (Supplementary Figure~S7), and a second region of the same sample shows the same behavior (Supporting Information, Section~S9), so the contrast reflects connectivity rather than proximity. Material near the electrode is therefore not necessarily part of the conduction path. This directly supports two simulation results. Current grows with connectivity rather than with the number of nodes (Figure~\ref{fig:N_sweep}), and introducing voids reduces the current even when the number of junctions and their activation voltages are unchanged (Figure~\ref{fig:void}), because material that is present but disconnected does not contribute to conduction. It also explains why networks of similar coverage can differ in threshold voltage~\cite{Wilson2019}.

Connectivity is not binary, however. Clusters separated from the connected network by gaps of a few nanometers appear brighter than those separated by larger gaps (Supporting Information, Section~S9), consistent with partial charge drainage by tunneling across the smallest gaps. The network thus forms a hierarchy of connectivity (Figure~\ref{fig:sem}e), in which a core connected directly to the electrode is surrounded by clusters that can reach it only across one or more nanometer-scale gaps in sequence. This hierarchy is the structural counterpart of two model features. The spread of gap widths is a plausible physical origin of the junction-to-junction spread in activation voltage, \(\sigma_{\mathrm{a}}\). The stepwise routes in Figure~\ref{fig:sem}e mirror the growth of the conducting subgraph above \(V_T\), as clusters behind the smallest gaps join first. Because the images are recorded at zero bias, this activation sequence is inferred rather than observed, and imaging under applied bias would allow it to be tested directly.

\subsection{Activation Disorder and Design Tradeoffs}

The preferred network structure depends on the intended application. Switching devices, selectors, and threshold sensors benefit from a rapid increase in current over a narrow voltage range, which provides a clear separation between off and on states~\cite{Sze2006, Waser2007, Wong2012, zhou2023thermally}. Analog sensing or neuromorphic applications may instead benefit from a gradual transition that provides intermediate conductance states~\cite{ding2022engineering}. Rather than having a single optimal structure, the network should be tuned through its junction characteristics and topology to balance the desired threshold voltage and switching sharpness.

\begin{figure*}[!hbt]
\centering
\includegraphics[width=.9\linewidth]{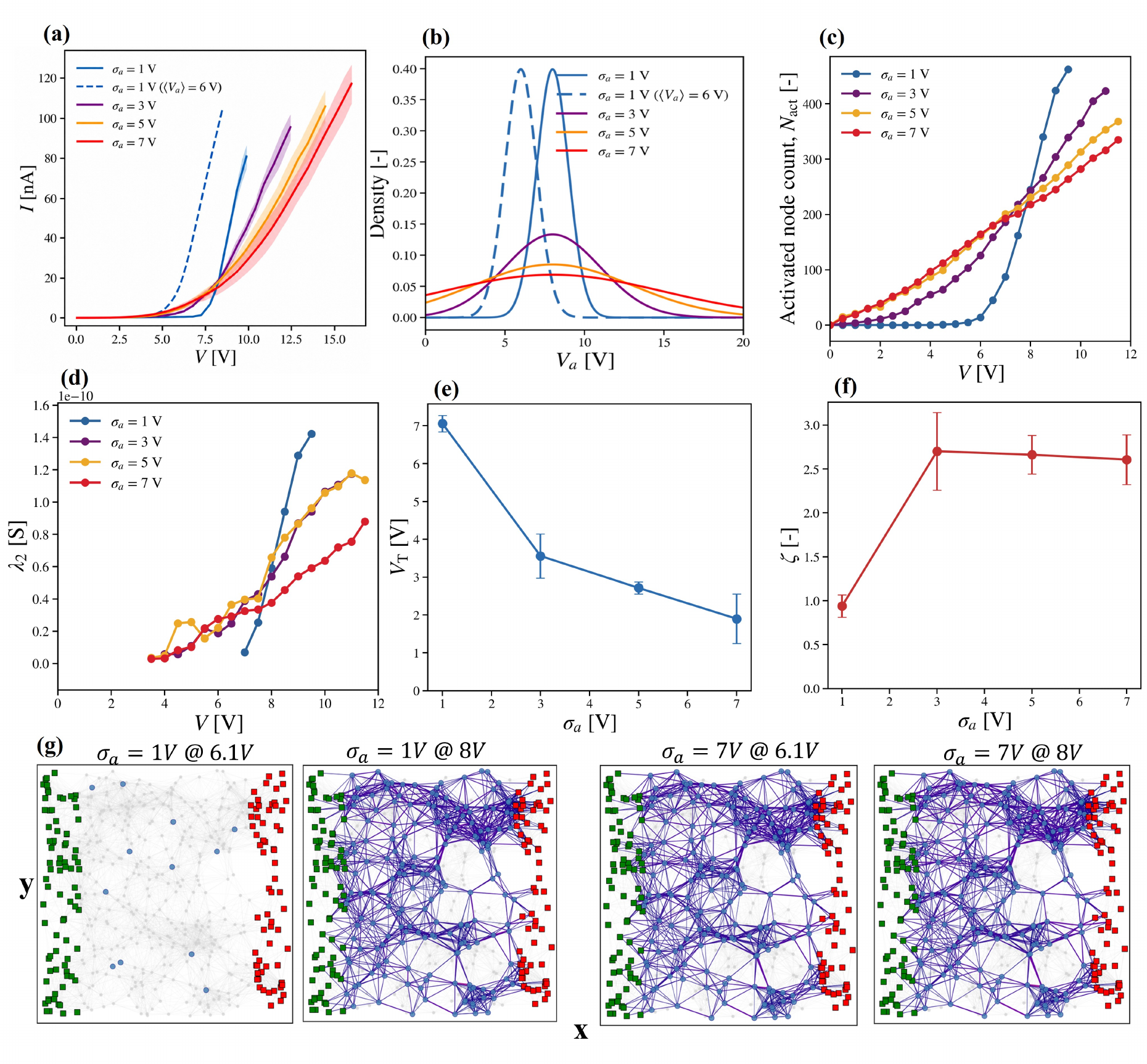}
\caption{Effect of activation-voltage disorder on conduction. (a)~\(I\)--\(V\) characteristics for \(\sigma_{\mathrm{a}}=1\), \(3\), \(5\), and \(7\)~V (\(\langle V_{\mathrm{a}}\rangle=8\)~V, \(N=500\), \(f_v=0.00\)); the dashed curve is a reference with \(\sigma_{\mathrm{a}}=1\)~V and \(\langle V_{\mathrm{a}}\rangle=6\)~V. The corresponding (b)~truncated activation-voltage distributions, (c)~activated-node count \(N_{\mathrm{act}}\), and (d)~conductance-weighted algebraic connectivity \(\lambda_2\). (e)~Threshold voltage \(V_T\) and (f)~scaling exponent \(\zeta\) versus \(\sigma_{\mathrm{a}}\), shown as means and standard deviations over five realizations. (g)~Active networks with \(\sigma_{\mathrm{a}}=1\) and \(7\)~V, each shown at \(6.1\) and \(8\)~V.}
\label{fig:design}
\end{figure*}

The width of the activation-voltage distribution, \(\sigma_{\mathrm{a}}\), controls how sharply the network turns on. A narrow distribution places most activation voltages close to the mean (Figure~\ref{fig:design}b), so few nodes activate at low bias and many then activate over a narrow voltage interval (Figure~\ref{fig:design}c). A broad distribution contains more nodes in both tails, so activation begins at lower bias, while high-\(V_{\mathrm{a}}\) nodes activate much later. The connectivity follows the same pattern. \(\lambda_2\) rises abruptly for small \(\sigma_{\mathrm{a}}\) but develops earlier and more gradually for large \(\sigma_{\mathrm{a}}\) (Figure~\ref{fig:design}d).The representative networks show this contrast directly (Figure~\ref{fig:design}g). Between \(6.1\) and \(8\)~V, the \(\sigma_{\mathrm{a}}=1\)~V network goes from nearly inactive to extensively connected, whereas the \(\sigma_{\mathrm{a}}=7\)~V network, already partly connected at \(6.1\)~V, changes only modestly. A narrow distribution therefore produces a later but abrupt turn-on, whereas a broad distribution produces an earlier but more gradual one.

These differences appear in both transport metrics.  \(V_T\) decreases as \(\sigma_{\mathrm{a}}\) increases (Figure~\ref{fig:design}e). A broad distribution provides more low-threshold nodes and thus a greater chance that an early source--drain path forms. \(\zeta\) decreases, reflecting the more gradual current rise above \(V_T\) (Figure~\ref{fig:design}f). The variation among network realizations also increases, because percolation becomes more sensitive to where these few early-activated nodes happen to lie. A low \(V_T\) obtained through large \(\sigma_{\mathrm{a}}\) therefore comes at the cost of a broader and less reproducible transition.

This result runs counter to a natural expectation. Because the nonlinear response of these networks arises from their disordered structure, one might expect greater junction-to-junction heterogeneity to produce a more pronounced non-Ohmic transition. The simulations show the opposite.  Heterogeneity spreads activation over a wider voltage range, whereas uniform junctions produce a steeper turn-on. The nonlinear $I$--$V$ response is not dominated by a few junctions with high conduction barriers, as might be assumed, but emerges collectively, with each junction contributing a comparable share. A sharp non-Ohmic response therefore requires homogeneous junction properties, with the collective transition supplied by how the junctions connect rather than by their disorder. For switching applications, fabrication should thus aim to minimize junction heterogeneity rather than rely on it.

The model further shows that threshold position and switching sharpness can be tuned separately. This matters because uniform junctions, while producing the sharpest turn-on, also delay it. The dashed reference curve in Figure~\ref{fig:design}a shows that reducing \(\langle V_{\mathrm{a}}\rangle\) while keeping \(\sigma_{\mathrm{a}}\) small shifts the steep \(I\)--\(V\) response to lower voltage without broadening it. \(\langle V_{\mathrm{a}}\rangle\) sets where the transition occurs, whereas \(\sigma_{\mathrm{a}}\) sets how widely the activation events are spread. A sharp and reproducible turn-on requires uniform junctions, sufficient network density, and a lower void fraction to avoid structural bottlenecks~\cite{Wilson2019}. In physical \(N^{3}\), both the mean and spread of the activation voltage are expected to depend on junction spacing, ligand chemistry, and the dielectric environment. Thus, junction uniformity controls switching sharpness, network topology controls pathway reliability, and the mean junction characteristics set the turn-on voltage.

\section{Conclusion}

We developed a graph-based Kirchhoff framework that connects the microscopic structure of self-assembled gold nanonecklace networks to their macroscopic electrical response. The model resolves the internal conduction of self-assembled gold nanonecklace networks, including where the first conducting path forms, how current redistributes, and how connectivity evolves with voltage. These internal states cannot be observed experimentally during operation, and the macroscopic response of such networks has so far been understood largely empirically. The model reproduces the experimentally observed \(I\)--\(V\) trends, including the shift of \(V_T\) with the junction activation scale, the lower threshold of denser networks, and the delayed percolation caused by voids. This agreement lends confidence to the internal network states that the model predicts.

The simulations show that junction properties and network topology control different features of the non-Ohmic response. The mean activation voltage \(\langle V_{\mathrm{a}}\rangle\), which reflects the typical junction spacing and local environment, shifts the conduction threshold without changing the shape of the transition. Network density instead reshapes the transition, with the current scaling approximately as \(I\propto N^{2.3}\). At the same time, the number of active nodes grows only linearly, so conduction is governed by how the junctions are connected rather than by how many are active. Voids weaken this connectivity and delay percolation even when the junction properties are unchanged.

These results also revise a natural expectation about these networks. Because their non-Ohmic behavior arises from a randomly connected structure, greater junction-to-junction heterogeneity might be expected to produce a stronger nonlinear response, with the junctions of highest conduction barrier dominating the transition. The simulations show the opposite. Heterogeneity lowers \(V_T\) but spreads activation over a wide voltage range, whereas homogeneous junctions combined with high connectivity produce the abrupt, reproducible turn-on required for sensors and transistor-like devices. The nonlinear response therefore emerges from the network as a whole rather than from a few high-barrier junctions. These findings turn the design of nanonecklace networks from empirical optimization into a rational choice of parameters, in which junction heterogeneity, network density, and void content are controlled during synthesis and fabrication to set the switching behavior, and the mean junction characteristics position the operating threshold.

\section*{Acknowledgments}
The authors acknowledge the Department of Chemical and Biomolecular Engineering at the University of Nebraska--Lincoln for support. The authors thank the Holland Computing Center at the University of Nebraska--Lincoln for providing computational resources.

\section*{Technology Use Disclosure}
During the preparation of this manuscript, the authors used ChatGPT (OpenAI) and Claude (Anthropic) to assist with grammar and typographical corrections. The authors reviewed and edited all content as needed and take full responsibility for the accuracy and integrity of the work.

\section*{Author Contributions}
O.I. developed and implemented the computational framework, performed the simulations and data analysis, prepared the figures, and drafted the manuscript. M.K.A.T. prepared the samples, performed the charge-contrast SEM imaging and image analysis, prepared the corresponding figures, and wrote the corresponding sections of the manuscript. S.B. contributed to model analysis, interpretation, and manuscript preparation. R.F.S. contributed to the experimental and physical context and the conceptual framework, and directed the charge-contrast SEM study. J.O. supervised the project and revised and edited the manuscript as the corresponding author. All authors reviewed and approved the final manuscript.

\section*{Conflicts of Interest}
The authors have no conflicts of interest to disclose.

\section*{Data Availability Statement}
The code and data can be found in the following publicly available GitHub repository: \url{https://github.com/ock-group/nanonet}. The SEM images and image-analysis scripts are available from the authors on request.

\section*{Supporting Information}
Supporting Information accompanies this article and contains graph construction and activation-voltage sampling (Section~S1), resistance-assignment details (Section~S2), Kirchhoff-solver validation and a worked three-node calculation (Section~S3), threshold and power-law fitting (Section~S4), algebraic-connectivity analysis (Section~S5), internal network diagnostics (Section~S6), structural-void construction (Section~S7), additional conducting-network snapshots (Section~S8; Supplementary Figures~S1--S6), and charge-contrast SEM controls and reproducibility (Section~S9; Supplementary Figures~S7 and S8).

\bibliography{reference}

@article{Lim2025,
   author = {Jay Min Lim and Muhammad Ashar Naveed and Yanan Wang and Ravi F. Saraf},
   doi = {10.1039/d4nr04770b},
   issn = {20403372},
   issue = {9},
   journal = {Nanoscale},
   month = {1},
   pages = {5012-5020},
   pmid = {39871599},
   publisher = {Royal Society of Chemistry},
   title = {Kinetics of ion-mediated directed self-assembly of one-dimensional chains of metal nanoparticles in solution},
   volume = {17},
   year = {2025}
}

@article{RevModPhys.79.469,
  title = {Granular electronic systems},
  author = {Beloborodov, I. S. and Lopatin, A. V. and Vinokur, V. M. and Efetov, K. B.},
  journal = {Rev. Mod. Phys.},
  volume = {79},
  issue = {2},
  pages = {469--518},
  numpages = {0},
  year = {2007},
  month = {Apr},
  publisher = {American Physical Society},
  doi = {10.1103/RevModPhys.79.469},
  url = {https://link.aps.org/doi/10.1103/RevModPhys.79.469}
}

@article{Wilson2019,
   author = {Peter Wilson and Jason K.Y. Ong and Abhijeet Prasad and Ravi F. Saraf},
   doi = {10.1021/acs.jpcc.9b05527},
   issn = {19327455},
   issue = {32},
   journal = {Journal of Physical Chemistry C},
   month = {8},
   pages = {19999-20005},
   publisher = {American Chemical Society},
   title = {Quantitative Visualization of Topology and Morphing of Percolation Path in Nanoparticle Network Array Exhibiting Coulomb Blockade at Room Temperature},
   volume = {123},
   year = {2019}
}

@article{Middleton1993,
  author  = {Middleton, A. Alan and Wingreen, Ned S.},
  title   = {Collective Transport in Arrays of Small Metallic Dots},
  journal = {Physical Review Letters},
  year    = {1993},
  volume  = {71},
  number  = {19},
  pages   = {3198--3201},
  doi     = {10.1103/PhysRevLett.71.3198}
}

@article{Prasad2021,
   author = {Abhijeet Prasad and Michael Stoller and Ravi F. Saraf},
   doi = {10.1021/acsanm.1c01641},
   issn = {25740970},
   issue = {9},
   journal = {ACS Applied Nano Materials},
   month = {9},
   pages = {9044-9051},
   publisher = {American Chemical Society},
   title = {Critical Behavior in Au Nanoparticle Arrays: Implications for All-Metal Field Effect Transistors with Ultra-high Gain at Room Temperature},
   volume = {4},
   year = {2021}
}

@article{Kane2010,
   author = {Jennifer Kane and Mehmet Inan and Ravi F. Saraf},
   doi = {10.1021/nn901161w},
   issn = {19360851},
   issue = {1},
   journal = {ACS Nano},
   month = {1},
   pages = {317-323},
   pmid = {20038126},
   title = {Self-assembled nanoparticle necklaces network showing single-electron switching at room temperature and biogating current by living microorganisms},
   volume = {4},
   year = {2010}
}

@article{Prasad2024,
   author = {Abhijeet Prasad and Jay Min Lim and Ravi F. Saraf},
   doi = {10.1002/aelm.202300485},
   issn = {2199160X},
   issue = {1},
   journal = {Advanced Electronic Materials},
   month = {1},
   publisher = {John Wiley and Sons Inc},
   title = {Conduction Mechanism Switching from Coulomb Blockade to Classical Critical Percolation Behavior in Disordered Nanoparticle Array},
   volume = {10},
   year = {2024}
}

@article{Elteto2004,
  author  = {Elteto, Klara and Antonyan, Eduard G. and Nguyen, T. T. and Jaeger, Heinrich M.},
  title   = {Model for the Onset of Transport in Systems with Distributed Thresholds for Conduction},
  journal = {Physical Review B},
  year    = {2005},
  volume  = {71},
  number  = {6},
  pages   = {064206},
  doi     = {10.1103/PhysRevB.71.064206}
}

@article{Parthasarathy2004,
  author  = {Parthasarathy, Raghuveer and Lin, Xiao-Min and Elteto, Klara and Rosenbaum, T. F. and Jaeger, Heinrich M.},
  title   = {Percolating through Networks of Random Thresholds: Finite-Temperature Electron Tunneling in Metal Nanocrystal Arrays},
  journal = {Physical Review Letters},
  year    = {2004},
  volume  = {92},
  number  = {7},
  pages   = {076801},
  doi     = {10.1103/PhysRevLett.92.076801}
}

@article{Blunt2007,
   author = {Matthew O. Blunt and Milovan {\v{S}}uvakov and Fabio Pulizzi and Christopher P. Martin and Emmanuelle Pauliac-Vaujour and Andrew Stannard and Andrew W. Rushforth and Bosiljka Tadi{\'c} and Philip Moriarty},
   doi = {10.1021/nl061656e},
   issn = {15306984},
   issue = {4},
   journal = {Nano Letters},
   month = {4},
   pages = {855-860},
   pmid = {17335264},
   title = {Charge transport in cellular nanoparticle networks: Meandering through nanoscale mazes},
   volume = {7},
   year = {2007}
}

@article{Parthasarathy2001,
   author = {Raghuveer Parthasarathy and Xiao Min Lin and Heinrich M. Jaeger},
   doi = {10.1103/PhysRevLett.87.186807},
   issn = {10797114},
   issue = {18},
   journal = {Physical Review Letters},
   pages = {186807-1-186807-4},
   title = {Electronic transport in metal nanocrystal arrays: The effect of structural disorder on scaling behavior},
   volume = {87},
   year = {2001}
}

@article{Reichhardt2003,
   author = {C. Reichhardt and C. J. Olson Reichhardt},
   doi = {10.1103/PhysRevLett.90.046802},
   issn = {10797114},
   issue = {4},
   journal = {Physical Review Letters},
   pages = {4},
   title = {Charge Transport Transitions and Scaling in Disordered Arrays of Metallic Dots},
   volume = {90},
   year = {2003}
}

@article{Tadic2015,
  author  = {Tadi{\'c}, Bosiljka and Andjelkovi{\'c}, Miroslav and {\v{S}}uvakov, Milovan},
  title   = {The Influence of Architecture on Collective Charge Transport in Nanoparticle Assemblies Revealed by the Fractal Time Series and Topology of Phase-Space Manifolds},
  journal = {Journal of Coupled Systems and Multiscale Dynamics},
  year    = {2016},
  volume  = {4},
  number  = {1},
  doi     = {10.1166/jcsmd.2016.1094}
}

@article{Ong2013,
    author = {Ong, Jason Kee Yang and Nguyen, Chieu Van and Sayood, Sena and Saraf, Ravi F.},
    title = {Imaging Electroluminescence from Individual Nanoparticles in an Array Exhibiting Room Temperature Single Electron Effect},
    journal = {ACS Nano},
    volume = {7},
    number = {8},
    pages = {7403-7410},
    year = {2013},
    month = {08},
    issn = {1936-0851},
    doi = {10.1021/nn403165q},
    url = {https://doi.org/10.1021/nn403165q},
    eprint = {https://pubs.acs.org/ancac3/article-pdf/7/8/7403/17961604/nn403165q.pdf},
}

@article{Modak2019,
    author = {Modak, Monami Das and Damarla, Ganesh and Maity, Somedutta and Chaudhary, Anil K. and Paik, Pradip},
    title = {Self-assembled pearl-necklace patterned upconverting nanocrystals with highly efficient blue and ultraviolet emission: femtosecond laser based upconversion properties},
    journal = {RSC Advances},
    volume = {9},
    number = {65},
    pages = {38246-38256},
    year = {2019},
    month = {11},
    issn = {2046-2069},
    doi = {10.1039/c9ra06389g},
    url = {https://doi.org/10.1039/c9ra06389g},
    eprint = {https://pubs.rsc.org/ra/article-pdf/9/65/38246/6860500/c9ra06389g.pdf},
}

@article{tran2008charge,
  title     = {Charge transport in self-assembled nanoparticle arrays: a review},
  author    = {Tran, T. B. and Beloborodov, I. S. and Lin, X. M. and Bigioni, T. P. and Jaeger, H. M.},
  journal   = {Annals of Physics},
  volume    = {323},
  number    = {10},
  pages     = {2324--2352},
  year      = {2008},
  publisher = {Elsevier}
}

@article{tadic2011modeling,
  title     = {Modeling collective charge transport in nanoparticle assemblies},
  author    = {Tadi{\'c}, Bosiljka and {\v{S}}uvakov, Milovan and Andjelkovi{\'c}, Miroslav},
  journal   = {Journal of Statistical Mechanics: Theory and Experiment},
  volume    = {2011},
  number    = {05},
  pages     = {P05014},
  year      = {2011},
  publisher = {IOP Publishing}
}

@article{Bascones2008,
   author = {E. Bascones and V. Est{\'e}vez and J. A. Trinidad and A. H. MacDonald},
   doi = {10.1103/PhysRevB.77.245422},
   issn = {10980121},
   issue = {24},
   journal = {Physical Review B - Condensed Matter and Materials Physics},
   month = {6},
   title = {Electronic correlations and disorder in transport through one-dimensional nanoparticle arrays},
   volume = {77},
   year = {2008}
}

@article{Narumi2011,
   author = {Takayuki Narumi and Masaru Suzuki and Yoshiki Hidaka and Shoichi Kai},
   doi = {10.1143/JPSJ.80.114704},
   issn = {00319015},
   issue = {11},
   journal = {Journal of the Physical Society of Japan},
   month = {11},
   title = {Size dependence of current-voltage properties in Coulomb blockade networks},
   volume = {80},
   year = {2011}
}

@article{Qu2017,
  author    = {Qu, Luman and V{\"o}r{\"o}s, M{\'a}rton and Zimanyi, Gergely T.},
  title     = {Metal-Insulator Transition in Nanoparticle Solids: Insights from Kinetic Monte Carlo Simulations},
  journal   = {Scientific Reports},
  year      = {2017},
  volume    = {7},
  number    = {1},
  pages     = {7071},
  month     = {Aug},
  doi       = {10.1038/s41598-017-06497-1},
  url       = {https://doi.org/10.1038/s41598-017-06497-1},
  issn      = {2045-2322}
}

@book{Sze2006,
  author    = {S. M. Sze and Kwok K. Ng},
  title     = {Physics of Semiconductor Devices},
  edition   = {3},
  publisher = {John Wiley \& Sons},
  year      = {2006}
}

@article{zhou2023thermally,
  author    = {Zhou, Xi and Zhao, Liang and Yan, Chu and Zhen, Weili and Lin, Yinyue and Li, Le and Du, Guanlin and Lu, Linfeng and Zhang, Shan-Ting and Lu, Zhichao and Li, Dongdong},
  title     = {Thermally stable threshold selector based on CuAg alloy for energy-efficient memory and neuromorphic computing applications},
  journal   = {Nature Communications},
  volume    = {14},
  number    = {1},
  pages     = {3285},
  year      = {2023},
  month     = {Jun},
  publisher = {Nature Publishing Group},
  doi       = {10.1038/s41467-023-39033-z},
  url       = {https://doi.org/10.1038/s41467-023-39033-z}
}

@article{Waser2007,
  author  = {Rainer Waser and Masakazu Aono},
  title   = {Nanoionics-Based Resistive Switching Memories},
  journal = {Nature Materials},
  volume  = {6},
  number  = {11},
  pages   = {833--840},
  year    = {2007},
  doi     = {10.1038/nmat2023}
}

@article{ding2022engineering,
  author    = {Ding, Ye and Zhang, Ying and Zhang, Xu and Chen, Peng and Zhang, Zhaokun and Yang, Yi and Cheng, Lin and Mu, Chengjun and Wang, Min and Xiang, Dong and Wu, Guangjun and Zhou, Kun and Yuan, Zengshuai and Liu, Qi},
  title     = {Engineering Spiking Neurons Using Threshold Switching Devices for High-Efficient Neuromorphic Computing},
  journal   = {Frontiers in Neuroscience},
  volume    = {15},
  pages     = {786694},
  year      = {2022},
  month     = {Jan},
  publisher = {Frontiers Media S.A.},
  doi       = {10.3389/fnins.2021.786694},
  url       = {https://doi.org}
}

@article{Wong2012,
  author  = {H.-S. Philip Wong and Sayeef Salahuddin},
  title   = {Memory Leads the Way to Better Computing},
  journal = {Nature Nanotechnology},
  volume  = {10},
  number  = {3},
  pages   = {191--194},
  year    = {2015},
  doi     = {10.1038/nnano.2015.29}
}

@article{Fiedler1973,
author = {Fiedler, Miroslav},
journal = {Czechoslovak Mathematical Journal},
language = {eng},
number = {2},
pages = {298-305},
publisher = {Institute of Mathematics, Academy of Sciences of the Czech Republic},
title = {Algebraic connectivity of graphs},
url = {http://eudml.org/doc/12723},
volume = {23},
year = {1973},
}

@article{Yu2013,
  author = {Yu, Chichao and Lee, Seung-Woo and Ong, Jason and Saraf, Ravi F.},
  title = {Single Electron Transistor in Aqueous Media},
  journal = {Advanced Materials},
  volume = {25},
  number = {22},
  pages = {3058-3062},
  year = {2013},
  doi = {10.1002/adma.201204162}
}

@article{Saraf2002,
  author  = {Saraf, Ravi F. and Lu, S. and Kindra, V. and Newbould, C.},
  title   = {Self-assembly of metal nanoparticles into organic-inorganic nanowires},
  journal = {Science},
  volume  = {296},
  number  = {5567},
  pages   = {510--513},
  year    = {2002},
  doi     = {10.1126/science.1070821}
}

@article{ZabetKhosousi2008,
  author  = {Zabet-Khosousi, Amir and Taghavi, Leila-Anis and Trudeau, Michel and Dhirani, Al-Amin},
  title   = {Charge Transport in Gapped Nanoparticle Arrays},
  journal = {ACS Nano},
  volume  = {2},
  number  = {3},
  pages   = {415--422},
  year    = {2008},
  doi     = {10.1021/nn700142q}
}

@article{Santhanam2016,
  author  = {Santhanam, Venugopal and Saraf, Ravi F.},
  title   = {Fabrication of high density, large-area nanoparticle arrays using ion-mediated self-assembly},
  journal = {Langmuir},
  volume  = {32},
  number  = {14},
  pages   = {3450--3457},
  year    = {2016},
  doi     = {10.1021/acs.langmuir.6b00347}
}

@article{Beloborodov2007,
  author  = {Beloborodov, I. S. and Lopatin, A. V. and Vinokur, V. M. and Efetov, K. B.},
  title   = {Granular electronic systems},
  journal = {Reviews of Modern Physics},
  volume  = {79},
  number  = {2},
  pages   = {469--518},
  year    = {2007},
  doi     = {10.1103/RevModPhys.79.469}
}

@article{Saraf2010,
  author  = {Saraf, Ravi F.},
  title   = {Self-Assembled Nanoparticle Arrays for Electronic and Electro-optical Applications},
  journal = {Chemical Reviews},
  volume  = {110},
  number  = {10},
  pages   = {5750--5788},
  year    = {2010},
  doi     = {10.1021/cr900137k}
}

@article{Cazaux2008,
   author = {Jacques Cazaux},
   title = {On some contrast reversals in {SEM}: Application to metal/insulator systems},
   journal = {Ultramicroscopy},
   volume = {108},
   issue = {12},
   pages = {1645-1652},
   year = {2008},
   doi = {10.1016/j.ultramic.2008.06.005}
}

@article{vanderWalt2014,
   author = {van der Walt, St{\'e}fan and Sch{\"o}nberger, Johannes L. and Nunez-Iglesias, Juan and Boulogne, Fran{\c{c}}ois and Warner, Joshua D. and Yager, Neil and Gouillart, Emmanuelle and Yu, Tony},
   title = {scikit-image: Image Processing in {Python}},
   journal = {PeerJ},
   volume = {2},
   pages = {e453},
   year = {2014},
   doi = {10.7717/peerj.453}
}

@inproceedings{Buades2005,
   author = {Buades, Antoni and Coll, Bartomeu and Morel, Jean-Michel},
   title = {A Non-Local Algorithm for Image Denoising},
   booktitle = {Proceedings of the 2005 IEEE Computer Society Conference on Computer Vision and Pattern Recognition (CVPR'05)},
   volume = {2},
   pages = {60-65},
   year = {2005},
   doi = {10.1109/CVPR.2005.38}
}

@article{Donoho1994,
   author = {Donoho, David L. and Johnstone, Iain M.},
   title = {Ideal Spatial Adaptation by Wavelet Shrinkage},
   journal = {Biometrika},
   volume = {81},
   number = {3},
   pages = {425-455},
   year = {1994},
   doi = {10.1093/biomet/81.3.425}
}

@article{Dijkstra1959,
   author = {Dijkstra, Edsger W.},
   title = {A Note on Two Problems in Connexion with Graphs},
   journal = {Numerische Mathematik},
   volume = {1},
   pages = {269-271},
   year = {1959},
   doi = {10.1007/BF01386390}
}
\end{document}


\tableofcontents
\newpage
\section{Graph Construction and Parameter Sampling}
\label{sec:SI_graph}

This section provides implementation details that supplement the graph definition and activation rule in the main text. Candidate nodes are sampled uniformly in a square domain of side length $L=1$. A spatial neighbor search identifies all pairs separated by less than $r_c=0.15L$. Nodes with degree zero or one are then removed iteratively. Because removing one node can reduce the degree of another, pruning continues until every retained internal graph node has a degree of at least two. Both the requested and realized node counts are written to the simulation output.

Source and drain contacts are assigned after pruning. Retained nodes with $x_i\leq0.15L$ belong to the source set, and nodes with $x_i\geq0.85L$ belong to the drain set. A realization is retained for electrical analysis only when both contact sets are nonempty.

The domain and connection radius remain fixed throughout the $N$-sweep. Thus, $N$ is a density-control parameter. Away from the domain boundaries, the expected degree is approximately

\begin{equation}
\langle k\rangle\approx\frac{N}{L^2}\pi r_c^2.
\label{eq:SI_expected_degree}
\end{equation}

For example, increasing $N$ from $200$ to $800$ at fixed $L$ and $r_c$ increases the expected interior degree by approximately a factor of four. The sweep therefore changes the number of nodes and the number of possible connections per node; it is not a pure system-size comparison.

Activation voltages are generated by rejection sampling. A candidate value is drawn from the specified normal distribution and accepted only when it lies within $0$--$20$~V. Sampling continues until every realized node has one valid value. This procedure preserves a genuinely truncated distribution and avoids artificial populations at the bounds that would result from clipping. The sampled values, random seed, requested parameters, and realized summary statistics are retained for reproducibility.

\section{Resistance-Assignment Details}
\label{sec:SI_resistance}

The main text defines the segment resistance, fixed node resistance, and symmetric total connection resistance. In the implementation, electrode-contact nodes have $\widetilde{R}_i=0$, whereas internal nodes have $\widetilde{R}_i=R_{\mathrm{node}}$. A numerical lower bound of $1~\Omega$ is applied after the total connection resistance is assembled. This floor is many orders of magnitude smaller than the simulated resistances and does not affect the reported results.

As a simple example, consider a series connection $n_1-n_2-n_3$, where $n_1$ and $n_3$ are electrode contacts and $n_2$ is an internal node. The two total connection resistances are

\begin{equation}
R_{12}^{\mathrm{tot}}
=R_{\mathrm{edge},12}+\frac{R_{\mathrm{node}}}{2},
\qquad
R_{23}^{\mathrm{tot}}
=R_{\mathrm{edge},23}+\frac{R_{\mathrm{node}}}{2}.
\label{eq:SI_series_resistance_example}
\end{equation}

Their sum contains exactly one complete internal-node resistance,

\begin{equation}
R_{12}^{\mathrm{tot}}+R_{23}^{\mathrm{tot}}
=R_{\mathrm{edge},12}+R_{\mathrm{node}}+R_{\mathrm{edge},23}.
\label{eq:SI_series_resistance_sum}
\end{equation}

This example shows how the symmetric half-node assignment reproduces the full resistance of an internal junction along an unbranched route without splitting the graph node into separate input and output terminals.

\section{Kirchhoff-Solver Details and Numerical Validation}
\label{sec:SI_kirchhoff_solver}

\subsubsection{Selection of the Conducting Subgraph}

After activation, connected components are identified by graph traversal. Two Boolean tests are applied to each component: whether it intersects the active source set and whether it intersects the active drain set. Components satisfying both tests are combined into the solved subgraph. All other components are removed before matrix assembly. This avoids adding electrically isolated matrix blocks that cannot contribute to the two-terminal current. If no component passes both tests, the voltage step is recorded as nonspanning and the electrical solve is skipped.

For example, an active cluster connected only to the source may contain many nodes and edges, but it has no return connection to the drain. Its internal potentials are therefore irrelevant to the measured source--drain current, and the cluster is excluded from the nodal system.

\subsubsection{Boundary-Current Calculation}

The current leaving the source electrode and entering the internal network is

\begin{equation}
J_{\mathrm{source}}
=
\sum_{s\in\mathcal{S}_V}
\sum_{\substack{j\notin\mathcal{S}_V\\(s,j)\in\mathcal{E}_V}}
g_{sj}(\phi_s-\phi_j),
\label{eq:SI_source_current}
\end{equation}

where \(\mathcal{S}_V\) is the set of active source-contact nodes and \(\mathcal{E}_V\) is the edge set of the spanning subgraph. Similarly, the current leaving the drain electrode and entering the network is

\begin{equation}
J_{\mathrm{drain}}
=
\sum_{d\in\mathcal{D}_V}
\sum_{\substack{j\notin\mathcal{D}_V\\(d,j)\in\mathcal{E}_V}}
g_{dj}(\phi_d-\phi_j),
\label{eq:SI_drain_current}
\end{equation}

where \(\mathcal{D}_V\) is the set of active drain-contact nodes. Edges connecting two nodes belonging to the same electrode are excluded because they do not cross the device boundary.

For a positive applied voltage, \(J_{\mathrm{source}}\) is normally positive because current leaves the source and enters the network. In contrast, \(J_{\mathrm{drain}}\) is normally negative because physical current leaves the network and enters the grounded drain. The corresponding nonnegative magnitudes are

\begin{equation}
I_{\mathrm{source}}=|J_{\mathrm{source}}|,
\qquad
I_{\mathrm{drain}}=|J_{\mathrm{drain}}|.
\label{eq:SI_boundary_magnitudes}
\end{equation}

The graph may store an undirected edge with either endpoint listed first. This storage order does not change the physical result. For each boundary contribution, the electrode node is placed first in the potential difference so that the current direction is defined consistently relative to that electrode.

\subsubsection{Current-Conservation Check}

At steady state, charge does not accumulate inside the network. Kirchhoff's current law therefore requires

\begin{equation}
J_{\mathrm{source}}+J_{\mathrm{drain}}=0,
\label{eq:SI_global_conservation}
\end{equation}

or, equivalently,

\begin{equation}
I_{\mathrm{source}}=I_{\mathrm{drain}}.
\label{eq:SI_boundary_equality}
\end{equation}

The reported current is

\begin{equation}
I(V)
=
\frac{
I_{\mathrm{source}}+I_{\mathrm{drain}}
}{2}.
\label{eq:SI_device_current}
\end{equation}

The symmetric average avoids choosing either electrode as the preferred current measurement. It is not used to correct a substantial current imbalance. In an accurate solution, the two boundary currents should already be equal apart from numerical roundoff.

The absolute current-balance error is recorded as

\begin{equation}
\Delta I_{\mathrm{bal}}
=
\left|
I_{\mathrm{source}}-I_{\mathrm{drain}}
\right|.
\label{eq:SI_balance_error}
\end{equation}

A negligible value of \(\Delta I_{\mathrm{bal}}\) confirms that the solved potentials and reconstructed edge currents satisfy global current conservation. A substantial value indicates a possible error in the matrix assembly, boundary-current calculation, or numerical solution.

\subsubsection{Worked Three-Node Example}

Consider the simplest conducting network containing three nodes arranged in series:

\[
n_1
\;-\;
n_2
\;-\;
n_3,
\]

where \(n_1\) is connected to the source, \(n_2\) is an internal node, and \(n_3\) is connected to the drain. The source and drain potentials are

\begin{equation}
\phi_1=V,
\qquad
\phi_3=0.
\label{eq:SI_three_node_boundaries}
\end{equation}

Let the conductances of edges \((1,2)\) and \((2,3)\) be

\begin{equation}
g_{12}=\frac{1}{R_{12}^{\mathrm{tot}}},
\qquad
g_{23}=\frac{1}{R_{23}^{\mathrm{tot}}}.
\label{eq:SI_three_node_conductances}
\end{equation}

The full conductance matrix is

\begin{equation}
\mathbf{G}
=
\begin{bmatrix}
g_{12} & -g_{12} & 0\\
-g_{12} & g_{12}+g_{23} & -g_{23}\\
0 & -g_{23} & g_{23}
\end{bmatrix}.
\label{eq:SI_three_node_matrix}
\end{equation}

Only \(\phi_2\) is unknown. Applying Kirchhoff's current law at \(n_2\) gives

\begin{equation}
g_{12}(\phi_2-\phi_1)
+
g_{23}(\phi_2-\phi_3)
=
0.
\label{eq:SI_three_node_KCL}
\end{equation}

Substituting \(\phi_1=V\) and \(\phi_3=0\) gives

\begin{equation}
(g_{12}+g_{23})\phi_2
=
g_{12}V.
\label{eq:SI_three_node_system}
\end{equation}

Therefore, the internal-node potential is

\begin{equation}
\phi_2
=
\frac{g_{12}}{g_{12}+g_{23}}V
=
\frac{R_{23}^{\mathrm{tot}}}
{R_{12}^{\mathrm{tot}}+R_{23}^{\mathrm{tot}}}V.
\label{eq:SI_three_node_potential}
\end{equation}

The current entering the network from the source is

\begin{equation}
J_{\mathrm{source}}
=
g_{12}(V-\phi_2),
\label{eq:SI_three_node_source}
\end{equation}

while the signed current leaving the drain and entering the network is

\begin{equation}
J_{\mathrm{drain}}
=
g_{23}(0-\phi_2).
\label{eq:SI_three_node_drain}
\end{equation}

Substituting Equation~\eqref{eq:SI_three_node_potential} gives

\begin{equation}
J_{\mathrm{source}}
=
\frac{V}
{R_{12}^{\mathrm{tot}}+R_{23}^{\mathrm{tot}}},
\qquad
J_{\mathrm{drain}}
=
-\frac{V}
{R_{12}^{\mathrm{tot}}+R_{23}^{\mathrm{tot}}}.
\label{eq:SI_three_node_boundary_results}
\end{equation}

Thus,

\begin{equation}
|J_{\mathrm{source}}|
=
|J_{\mathrm{drain}}|
=
I(V),
\label{eq:SI_three_node_conservation}
\end{equation}

and the device current is

\begin{equation}
I(V)
=
\frac{V}
{R_{12}^{\mathrm{tot}}+R_{23}^{\mathrm{tot}}}.
\label{eq:SI_three_node_device_current}
\end{equation}

The effective conductance of this three-node series network is therefore

\begin{equation}
G_{\mathrm{eff}}
=
\frac{1}
{R_{12}^{\mathrm{tot}}+R_{23}^{\mathrm{tot}}}.
\label{eq:SI_three_node_effective_conductance}
\end{equation}

This example shows the complete calculation sequence: impose the electrode potentials, solve the internal-node potential, calculate the edge currents, and sum the boundary currents. The same procedure is applied to the much larger active nanonecklace graph using a sparse linear solver.

\subsubsection{Numerical Safeguards}

For \(V>10^{-12}\)~V, the effective conductance is calculated as

\begin{equation}
G_{\mathrm{eff}}(V)=\frac{I(V)}{V}.
\label{eq:SI_effective_conductance}
\end{equation}

At \(V=0\), it is assigned a value of zero to avoid the undefined ratio \(0/0\). The \(10^{-12}\)-V cutoff is only a numerical safeguard and is not a physical transport threshold.

Roundoff-level edge currents are excluded from current-distribution statistics and visualizations. An edge is classified as current carrying when

\begin{equation}
|I_{ij}|
>
\max\!\left[
10^{-30}\ \mathrm{A},
10^{-12}
\max_{(k,l)\in\mathcal{E}_V}|I_{kl}|
\right].
\label{eq:SI_current_tolerance}
\end{equation}

The first term provides an absolute numerical floor, and the second provides a tolerance relative to the largest edge current at that voltage. This filtering is used only to identify and visualize current-carrying edges. It does not change the Kirchhoff solution or the calculated device current.

All reported potentials, edge currents, boundary currents, and current-distribution statistics are reconstructed from the same Kirchhoff solution. No separate path-current approximation or summation over simple paths is performed.
\section{Threshold and Power-Law Analysis}
\label{sec:SI_threshold}

The threshold and fitting window are defined in the main text. Computationally, the voltage sweep is searched in ascending order, and the first sampled point having both a spanning state and positive device current is stored as $V_T\equiv V_{\mathrm{perc}}$. The uncertainty associated with this discrete identification is limited by the voltage increment; for a $0.5$-V sweep, the onset is resolved only to the sampled $0.5$-V grid.

The threshold point itself is not included in the logarithmic fit because $V-V_T=0$. Positive-current points above $V_T$ are retained until, but not including, the first voltage at which the active-node fraction reaches $0.90$. The retained points are fitted by linear least squares using
\begin{equation}
  \ln I=\ln A+\zeta\ln(V-V_T).
  \label{eq:log_powerlaw_fit}
\end{equation}

For example, if the first nonzero spanning state occurs at $4.0$~V and $90\%$ activation is first reached at $12.0$~V, a sweep with $0.5$-V spacing uses the positive-current points from $4.5$ through $11.5$~V. The value $V_T=4.0$~V is held fixed, and the slope of $\ln I$ against $\ln(V-V_T)$ gives $\zeta$. The $90\%$ activation voltage is only the upper fitting boundary and is not a second threshold.

\section{Algebraic Connectivity}
\label{sec:SI_connectivity}

Connectivity is quantified from the same conductance-weighted Laplacian used by the Kirchhoff solver. For a connected graph, the algebraic connectivity is the second-smallest eigenvalue, $\lambda_2$, of this Laplacian~\cite{Fiedler1973}. A small positive $\lambda_2$ indicates a weakly connected structure containing bottlenecks, whereas a larger value indicates a more strongly interconnected conducting network.

Before applying the Dirichlet boundary conditions, the active-edge conductances are assembled into the symmetric matrix $\mathbf{G}$ defined in the main text. Its eigenvalues are evaluated with a symmetric eigensolver. To prevent physically meaningful but small conductance eigenvalues from being classified as numerical zero, the tolerance is scaled by the largest eigenvalue:
\begin{equation}
  \tau=10^{-9}\lambda_{\max}.
  \label{eq:eig_tol}
\end{equation}
Eigenvalues smaller than $\tau$ are treated as numerical zero modes, and $\lambda_2$ is reported as the smallest eigenvalue above this tolerance for the connected spanning network. Before percolation, no spanning circuit exists and $\lambda_2$ is reported as undefined. Because the Laplacian is weighted by electrical conductance, both $\lambda_2$ and $\lambda_{\max}$ have units of siemens. The ratio $\lambda_2/\lambda_{\max}$ is dimensionless.

\section{Internal Network Diagnostics}
\label{sec:SI_diagnostics}

At every voltage, the model records the numbers of activated nodes and edges, nodes and edges in the solved spanning network, connected components, and current-carrying edges. The sizes of the largest and second-largest activated components, $S_1$ and $S_2$, describe cluster growth before percolation, while the binary spanning state identifies the appearance of a continuous source--drain connection.

Current redistribution is quantified using the current participation ratio
\begin{equation}
  N_{\mathrm{eff}}
  =\frac{\left(\sum_e|I_e|\right)^2}
  {\sum_e|I_e|^2},
  \label{eq:participation_ratio}
\end{equation}
which estimates the effective number of edges sharing the current. A small $N_{\mathrm{eff}}$ indicates localization on a few edges, whereas a large value indicates broadly distributed transport. The model also records the coefficient of variation, Gini coefficient, maximum-to-mean current ratio, and fraction of total edge-current magnitude carried by the largest $10\%$ of current-carrying edges. Backbone edges are defined as edges satisfying $|I_e|\geq0.01\max|I_e|$ at that voltage. The effective source--drain resistance is obtained from a unit-bias solution using the same activated topology.

At $V_T=V_{\mathrm{perc}}$, each current-diagnostic figure pairs a histogram of $|I_{ij}|$ with the corresponding spatial current snapshot. Both panels use exactly the same current-carrying edge set defined by Equation~\eqref{eq:SI_current_tolerance}. Consequently, the sum of all histogram-bin counts equals the number of colored current-carrying edges in the snapshot. The height of the tallest histogram bin is only the number of edges within that current interval and is not the total number of current-carrying edges. The annotation reports $N_{\mathrm{act}}$, the number of active nodes at onset, together with the threshold voltage and global current. When a panel summarizes several realizations, these scalar annotations may be ensemble means and standard deviations, whereas the displayed spatial network is one representative realization. Within each realization, however, the histogram and snapshot use the same edge set. The per-voltage and per-edge CSV files retain the values needed to reconstruct both panels from the same Kirchhoff solution.

\section{Structural Voids}
\label{sec:SI_voids}

Void centers are sampled uniformly within the interior interval $[r_v+0.02L,L-r_v-0.02L]$ in each coordinate, ensuring that each complete circle remains inside the modeled domain with an additional boundary margin. Candidate graph nodes falling inside any void are rejected. After node placement, a proposed edge is also rejected when its line segment intersects a void, even when both endpoint nodes lie outside that void.

Because the circular regions may overlap, summing their individual areas can overestimate the physically excluded area. The requested fraction is therefore stored separately from the achieved union-area fraction. The latter is calculated on a deterministic $700\times700$ spatial grid so that an overlapping region is counted only once. For example, if two void circles overlap, the shared lens-shaped area contributes once to the achieved fraction rather than once for each circle. The raw parameter sweep includes $f_v=0.00$--$0.25$, while $f_v=0.25$ is excluded from the final publication figures.

\section{Conducting-Network Snapshots}
\label{sec:SI_snapshots}

The supplementary snapshots show how the conducting network changes with applied voltage. Before $V_T$, active nodes form isolated clusters that do not connect the electrodes. At $V_T$, the first current-carrying source--drain backbone appears. Further activation adds branches and parallel routes, redistributing current and strengthening the spanning network. At high voltage, activation approaches saturation and the conducting topology changes only weakly.

\begin{figure*}[p]
\centering
\includegraphics[width=\linewidth]{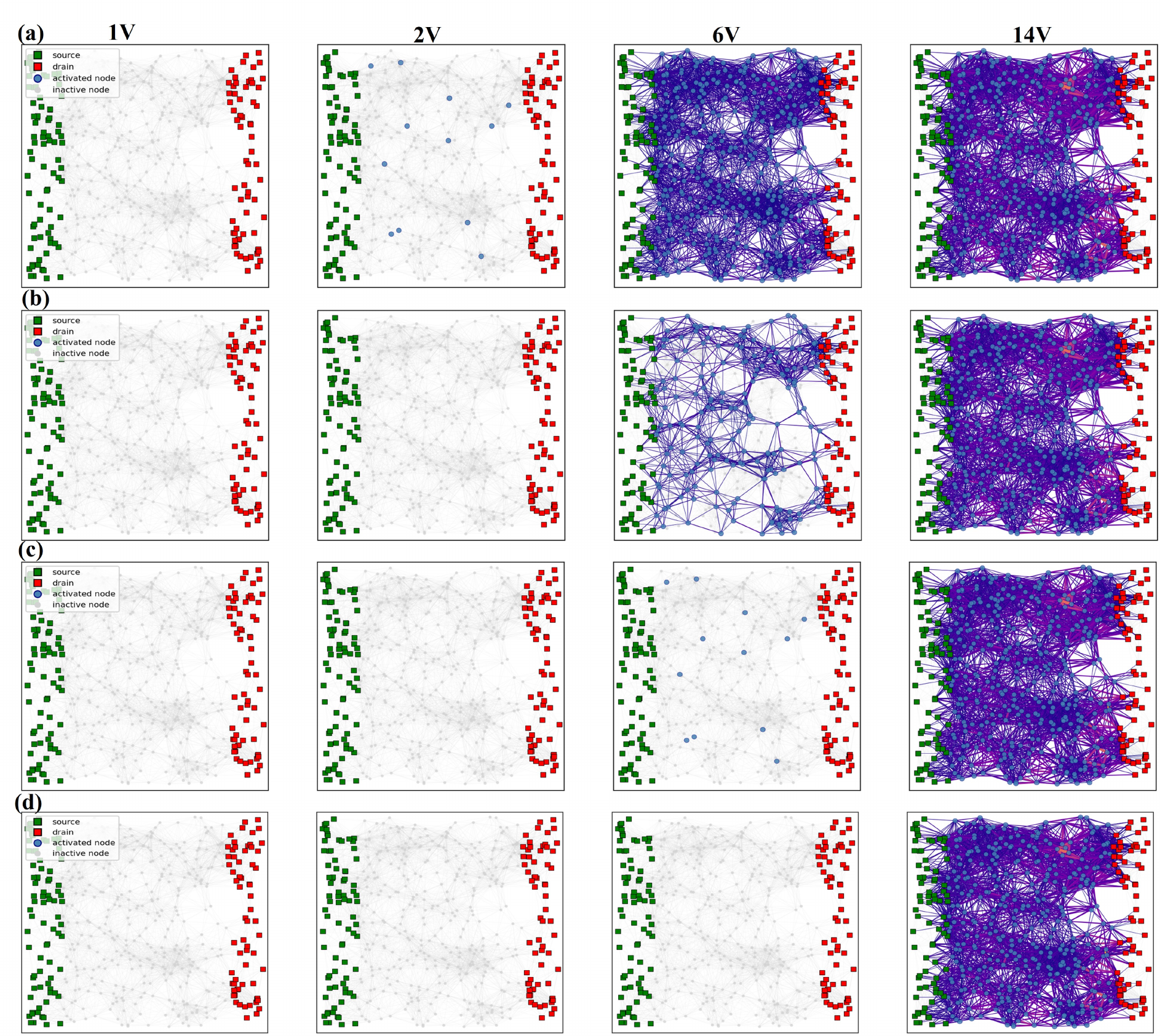}
\caption{\textbf{Effect of mean activation voltage on voltage-dependent network activation.}
Snapshots are shown for $N=500$ and $\sigma_a=1$~V at applied voltages of 1, 2, 6, and 14~V. Panels (a)--(d) correspond to $\langle V_{\mathrm{a}}\rangle=4$, 6, 8, and 10~V, respectively. Source and drain contacts are shown in green and red. Activated nodes are highlighted in blue, activated connections are colored, and inactive network elements are shown in light gray. Increasing $\langle V_{\mathrm{a}}\rangle$ delays activation to higher applied voltage.}
\label{fig:S1}
\end{figure*}

\begin{figure*}[p]
\centering
\includegraphics[width=\linewidth]{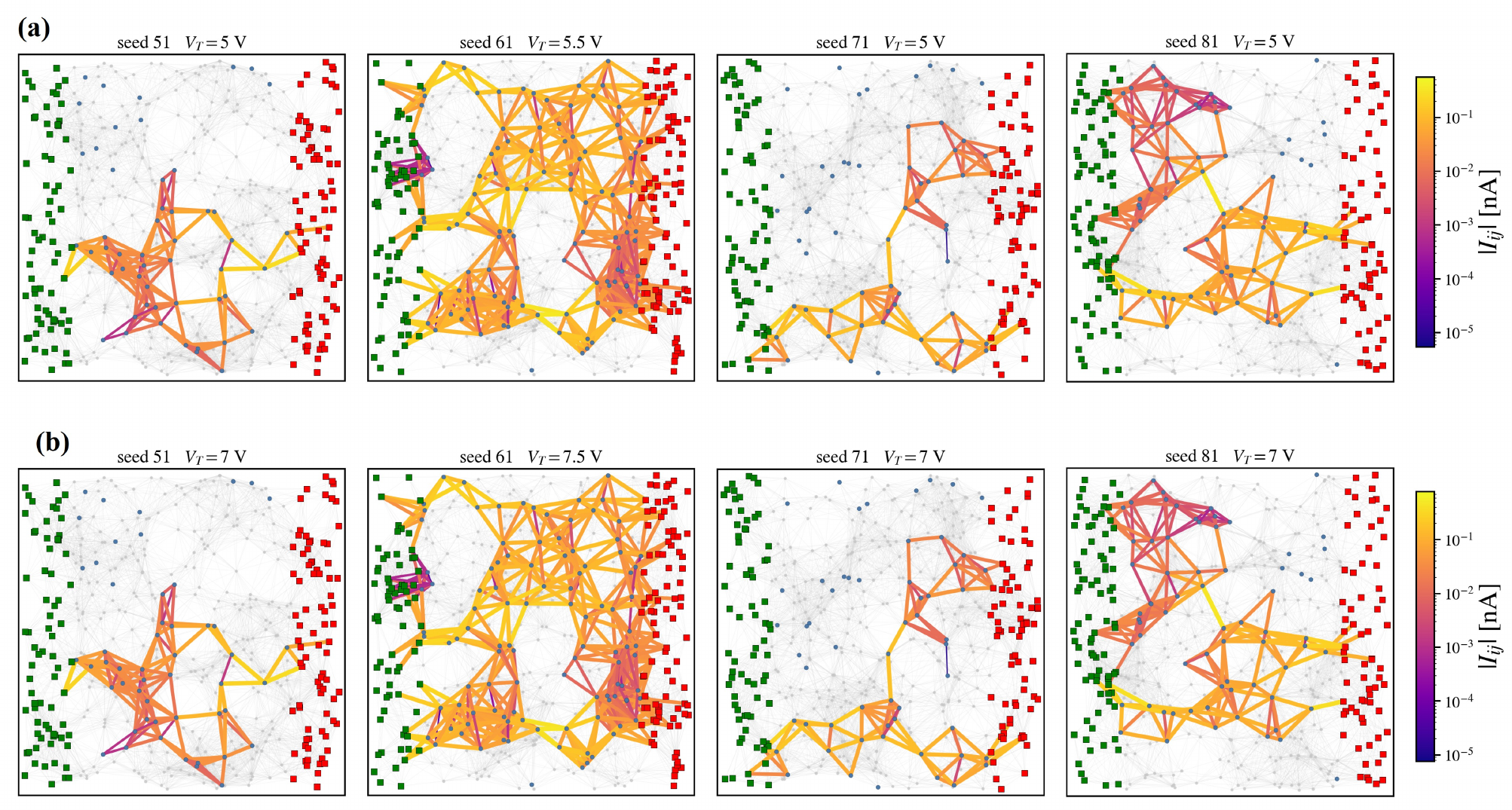}
\caption{\textbf{Seed-to-seed variability of threshold current pathways at two mean activation voltages.}
Panels (a) and (b) show $\langle V_{\mathrm{a}}\rangle=6$ and 8~V, respectively, for seeds 51, 61, 71, and 81. Each realization is displayed at its own threshold voltage $V_T$, which is annotated above the corresponding network. Source and drain contacts are shown in green and red, and current-carrying edges are colored by the magnitude $|I_{ij}|$ using the adjacent logarithmic color scale. The comparison illustrates that different random realizations can produce distinct conducting backbones even at the same nominal parameter values.}
\label{fig:S2}
\end{figure*}

\begin{figure*}[p]
\centering
\includegraphics[width=\linewidth]{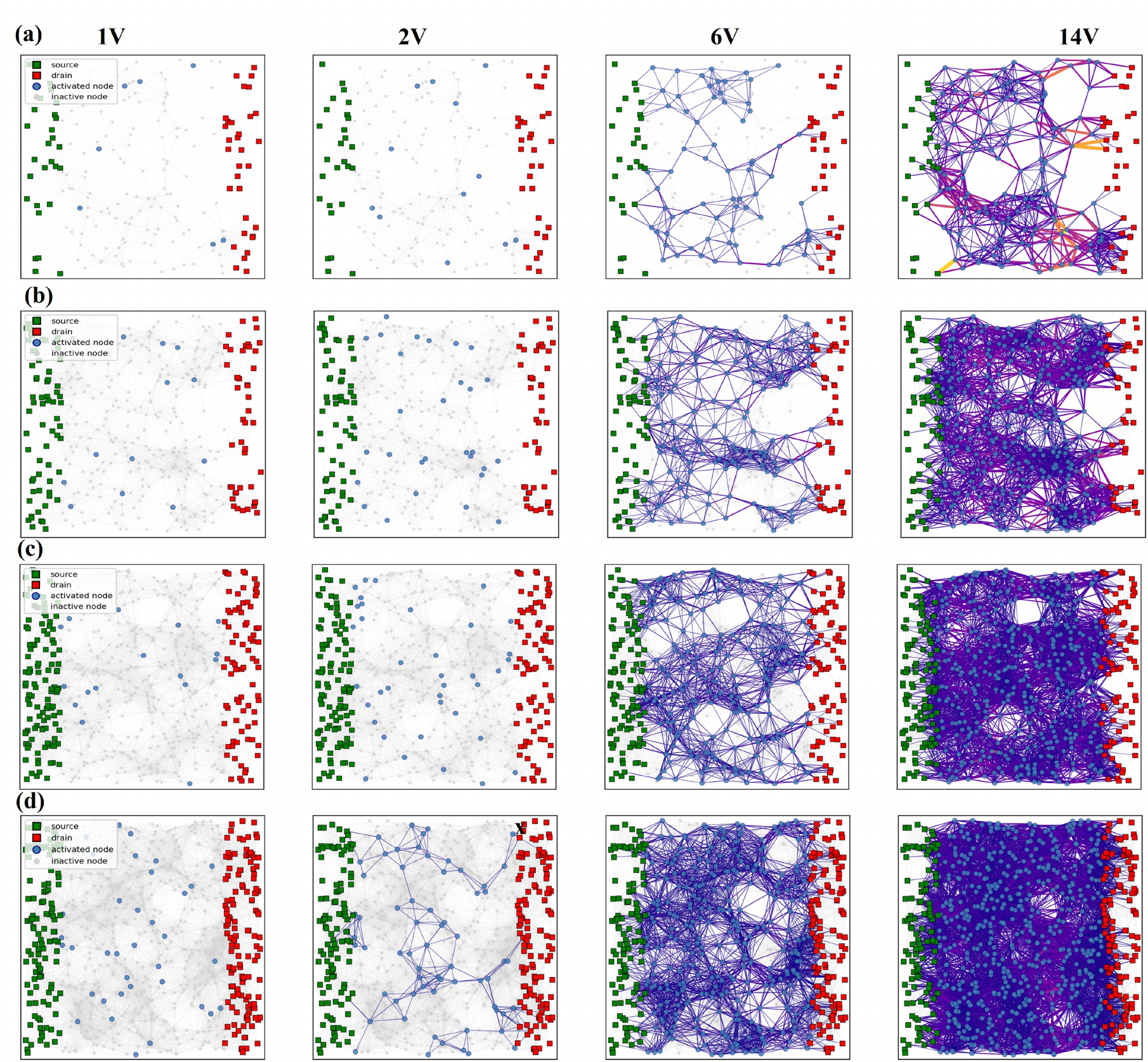}
\caption{\textbf{Voltage-dependent connectivity evolution with increasing network density.}
Panels (a)--(d) correspond to $N=200$, 400, 600, and 800 at fixed domain size, connection radius, $\langle V_{\mathrm{a}}\rangle=6$~V, and $\sigma_a=3$~V. Each panel shows the network at applied voltages of 1, 2, 6, and 14~V. Because the domain and connection radius are fixed, increasing $N$ increases the density of available nodes and connections.}
\label{fig:S3}
\end{figure*}

\begin{figure*}[p]
\centering
\includegraphics[width=\linewidth]{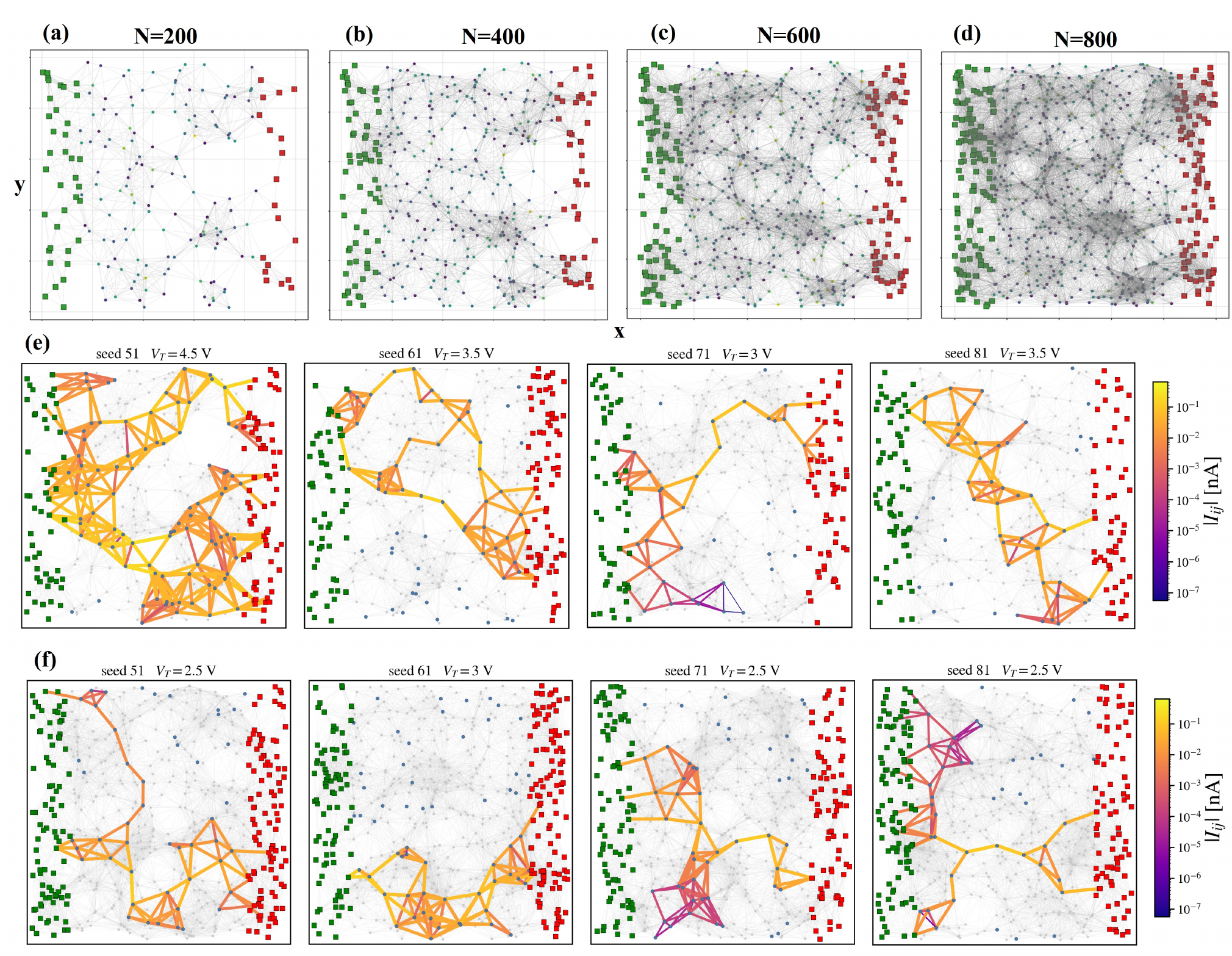}
\caption{\textbf{Network-density comparison and seed-to-seed threshold-current variability.}
Panels (a)--(d) show representative network realizations for $N=200$, 400, 600, and 800. Panel (e) shows threshold-current pathways for $N=400$ and panel (f) shows the corresponding pathways for $N=600$; within each of panels (e) and (f), seeds 51, 61, 71, and 81 are displayed from left to right. The threshold voltage $V_T$ is annotated for each realization, and current-carrying edges are colored by $|I_{ij}|$. The density study uses fixed $\langle V_{\mathrm{a}}\rangle=6$~V, $\sigma_a=3$~V, domain size, and connection radius.}
\label{fig:S4}
\end{figure*}

\begin{figure*}[p]
\centering
\includegraphics[width=\linewidth]{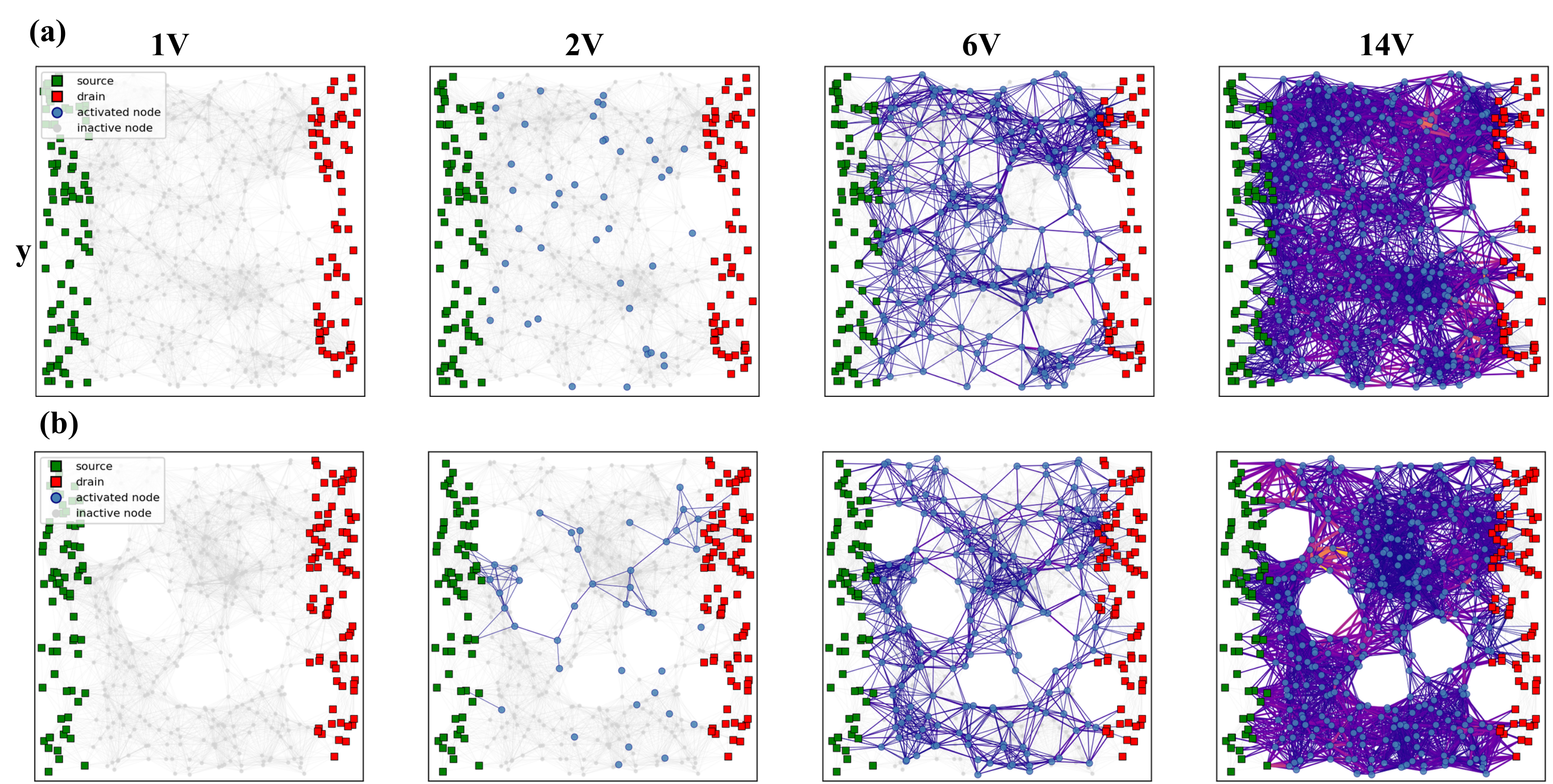}
\par\medskip
\includegraphics[width=\linewidth]{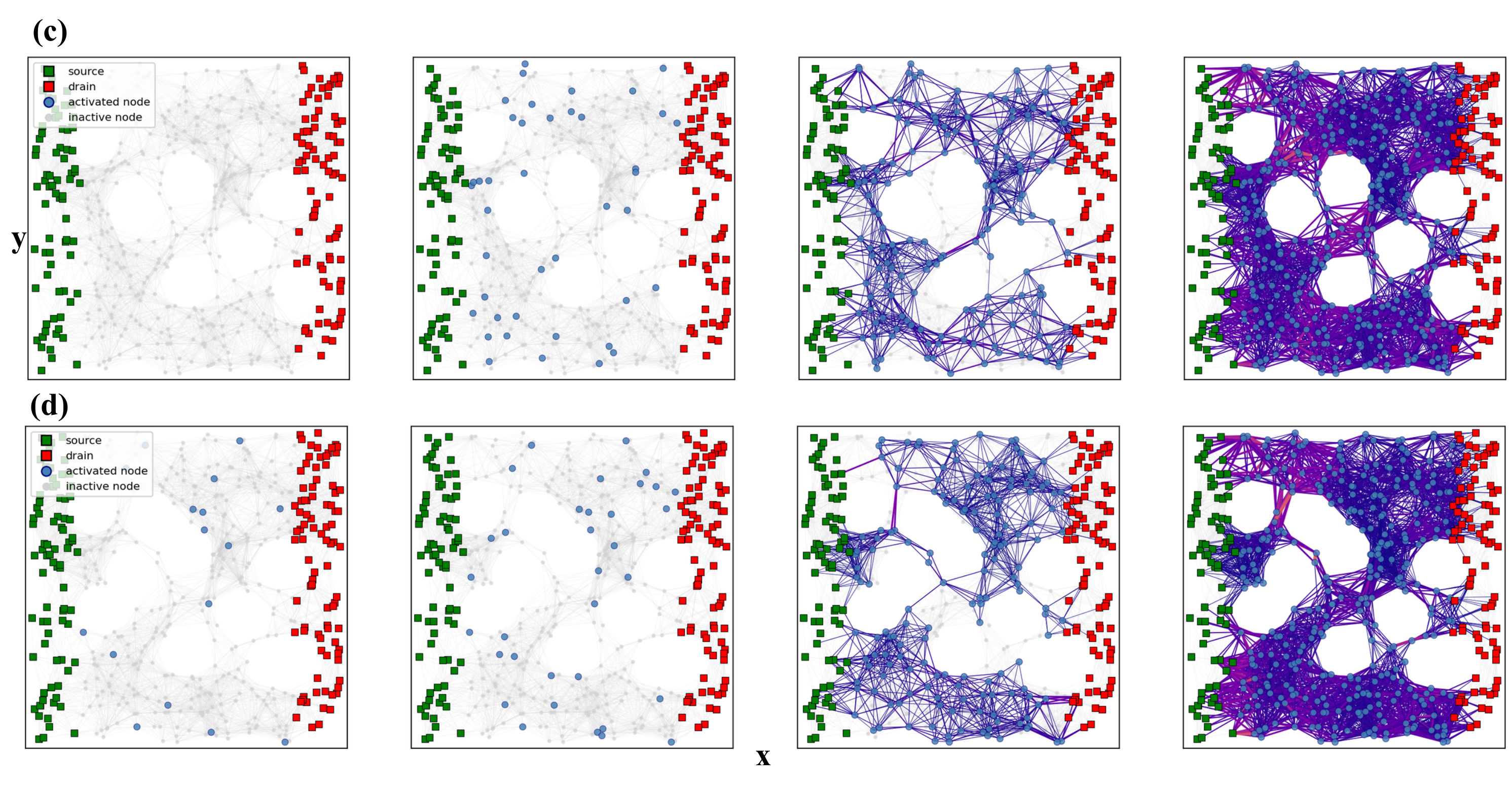}
\caption{\textbf{Influence of structural voids on voltage-dependent network activation.}
Snapshots are shown for $N=500$, $\langle V_{\mathrm{a}}\rangle=6$~V, and $\sigma_a=3$~V at (a) $f_v=0.00$, (b) $f_v=0.10$, (c) $f_v=0.15$, and (d) $f_v=0.20$. Each panel shows applied voltages of 1, 2, 6, and 14~V. Increasing void fraction removes available regions of the network and constrains the routes through which a source--drain connection can develop. The parameter sweep also contains $f_v=0.05$; $f_v=0.25$ is excluded from the final publication figures.}
\label{fig:S5}
\end{figure*}

\begin{figure*}[p]
\centering
\includegraphics[width=\linewidth]{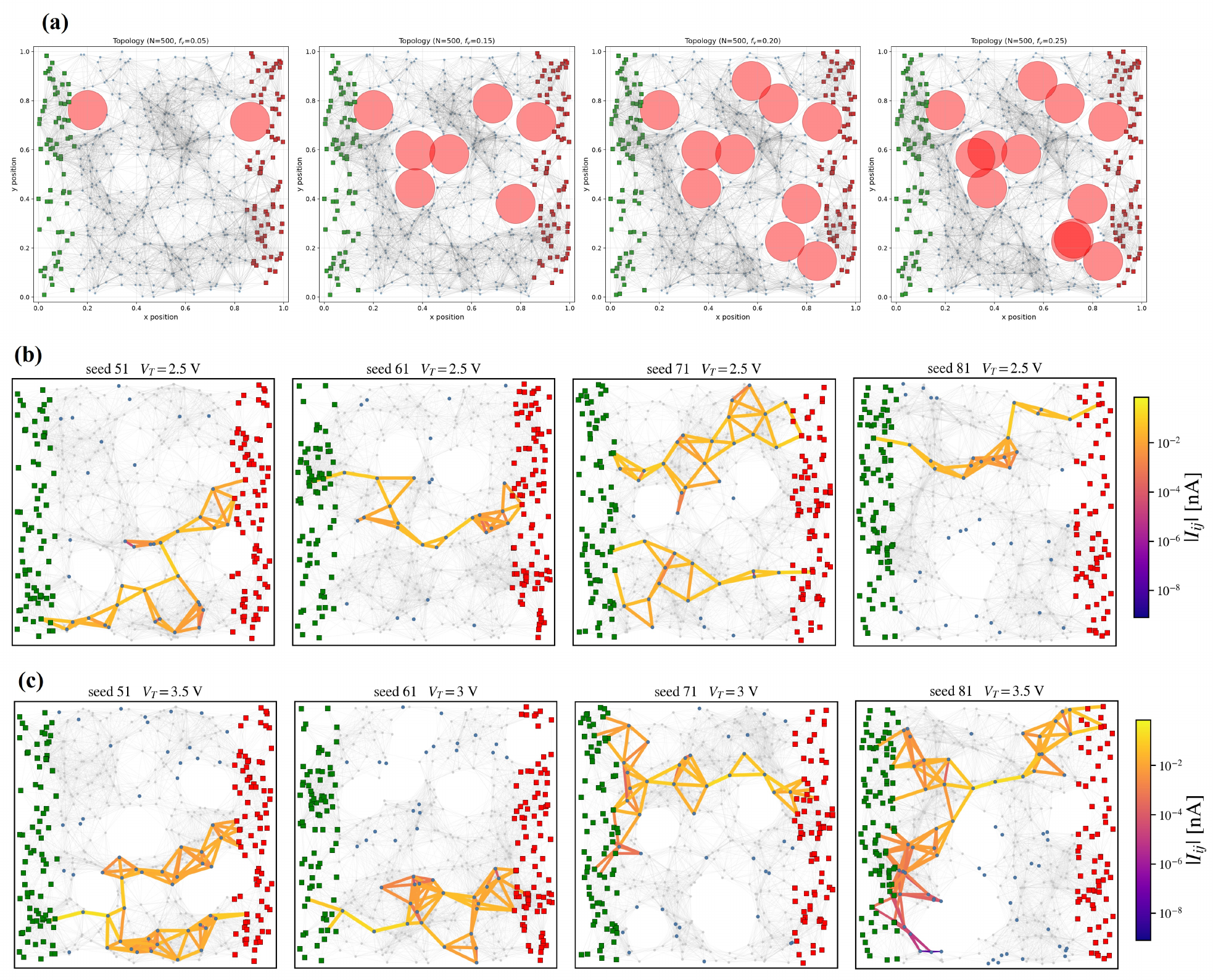}
\caption{\textbf{Seed-to-seed variability of threshold current pathways in networks with structural voids.}
Panel (a) shows the void placement in the necklace structure; panels (b) and (c) correspond to $f_v=0.05$ and 0.10, respectively, for $N=500$, $\langle V_{\mathrm{a}}\rangle=6$~V, and $\sigma_a=3$~V. Within each panel, seeds 51, 61, 71, and 81 are shown from left to right. Each realization is displayed at its annotated threshold voltage $V_T$. Source and drain contacts are shown in green and red, and current-carrying edges are colored by $|I_{ij}|$ using the adjacent logarithmic color scale.}
\label{fig:S6}
\end{figure*}

Supplementary Figure~\ref{fig:S1} shows the voltage-dependent activation sequence as $\langle V_{\mathrm{a}}\rangle$ is increased. Supplementary Figure~\ref{fig:S2} complements this comparison with seed-to-seed current-pathway maps at the threshold voltage. Supplementary Figure~\ref{fig:S3} shows the corresponding voltage-dependent evolution as the network density is increased through $N$, while Supplementary Figure~\ref{fig:S4} summarizes the associated topology and threshold-current variability. Supplementary Figure~\ref{fig:S5} shows the voltage-dependent evolution with increasing structural void fraction, and Supplementary Figure~\ref{fig:S6} shows representative threshold-current pathways for two nonzero void fractions.

\FloatBarrier
\newpage
\section{Charge-Contrast SEM: Controls and \mbox{Reproducibility}}

\textit{Analysis details.} Intensities are 16-bit values scaled to 0 to 1. The noise level \(\sigma\) was estimated with the wavelet-based estimator in scikit-image; non-local means used \(5\times5\) pixel patches and a search distance of \(6\) pixels (\(7\) pixels for the displayed images). The background was a grey-level opening (\(12\)-pixel disk) of the image smoothed with a Gaussian filter of \(1.5\) pixels standard deviation, followed by Gaussian smoothing with a standard deviation of \(10\) pixels. Chain pixels lie more than \(0.07\) above this background or are saturated; the mask was opened with a one-pixel disk, objects of \(60\) pixels or fewer and holes of \(40\) pixels or fewer were removed, and the mask was smoothed (Gaussian filter, standard deviation \(1.5\) pixels) and thresholded at \(0.5\). In each row, the sidewall foot was placed at the strongest rise in chain-excluded substrate brightness within \(45\) pixels of a reference line, smoothed along the edge and held within \(25\) pixels of a straight fit. Distance from the electrode is the horizontal distance to this line. A connected reference required at least \(300\) pixels. Gaps are the shortest pixel-centre distance between components minus one pixel; junctions in Figure~6e of the main text are gaps of \(10\)~nm or less. Routes in Figure~6e used 8-neighbour steps (MCP\_Geometric in scikit-image) with a cost of \(1\) per chain pixel and \(400\) per gap pixel; bare substrate could not be crossed. Raising the gap cost to \(10^{7}\) left the number of junctions unchanged for all \(69\) reachable clusters. The SEM images and analysis scripts reproduce the values and figures in this section.

\textit{Controls.} Brightness increases with local chain density for both classes, and the connected network is denser. When each isolated cluster is compared with connected chains at the same distance and the same local density (fraction of a disk of \(20\)~nm radius covered by chains, matched within \(\pm0.05\)), \(103\) of \(122\) clusters remain greyer in the main region and \(86\) of \(113\) in the second region, so connectivity contributes independently of density. Repeating the analysis on a more strongly denoised image (\(h=1.4\sigma\)) gives \(89\) of \(94\) clusters greyer within \(0.6\)~\(\mu\)m of the electrode.

\textit{Brightness by gap.} Grouped by gap to the connected network, the median isolated-to-connected brightness ratio in the main region is \(0.75\) below \(4\)~nm (\(15\) of \(18\) clusters greyer), \(0.55\) for \(4\) to \(10\)~nm (\(13\) of \(14\)), and \(0.45\) for larger gaps (\(120\) of \(122\)). The ratios are recorded intensities, so their absolute values depend on the detector offset; their order does not.

\textit{Second region.} A second region of the same sample (\(0.2\)~nA, \(10\)~\(\mu\)s, \(1.35\)~nm per pixel; Figure~\ref{fig:S8}) gives the same result: \(61\%\) of the chain area is isolated, and within \(1\)~\(\mu\)m of the electrode \(102\) of \(114\) isolated clusters are greyer than connected chains at the same distance. Referenced to the bare substrate, which is fully recorded in this region, isolated clusters carry a median of \(62\%\) of the signal of connected chains.

\begin{figure*}[p]
\centering
\includegraphics[width=\linewidth]{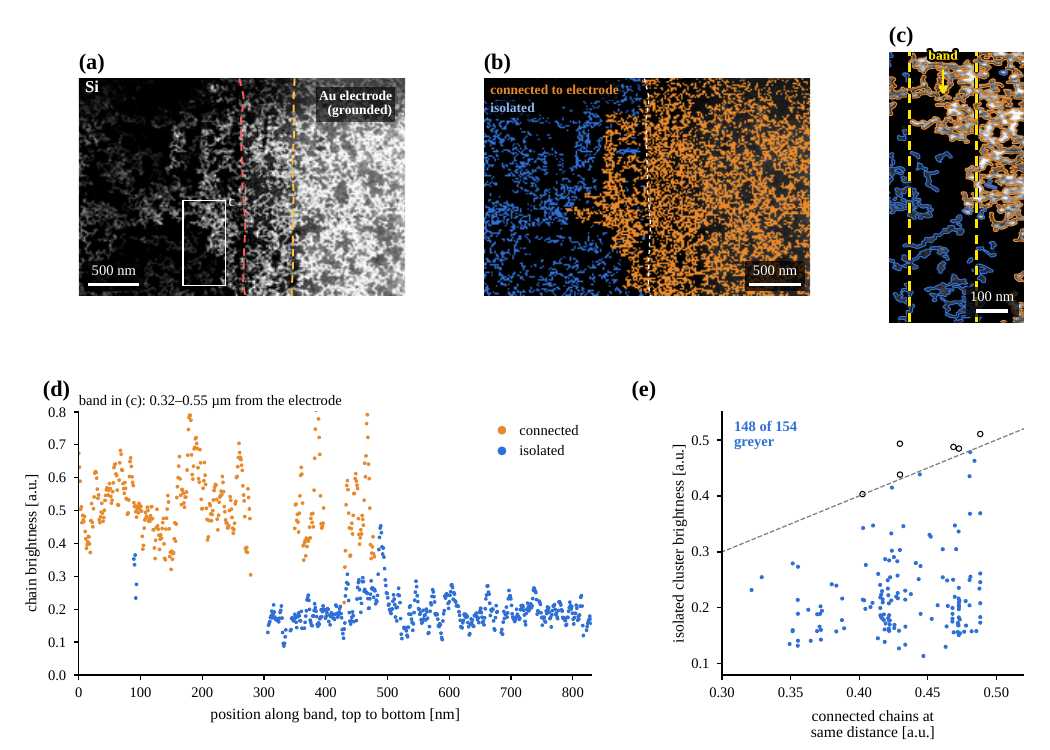}
\caption{\textbf{Connected chains are brighter than isolated clusters at the same distance from the electrode.}
(a)~Main region (\(5\)~kV, \(1.04\)~nm per pixel). Dashed lines mark the foot (red) and top (yellow) of the electrode sidewall; the box marks (c). (b)~Connected chains (orange) and isolated clusters (blue). (c)~Enlarged region (box in a). The yellow dashed lines bound a band \(0.32\) to \(0.55\)~\(\mu\)m from the electrode, parallel to its edge, so all chains in it lie at the same distance from the electrode; the arrow gives the direction of the position axis in (d). (d)~Mean chain brightness along this band, from top to bottom of (c): brightness drops sharply where the connected network (orange) ends and isolated clusters (blue) begin, although the distance to the electrode does not change. (e)~Brightness of each isolated cluster (at least \(150\) pixels) against connected chains at the same distance; \(148\) of \(154\) lie below the line of equal brightness (open circles, exceptions).}
\label{fig:S7}
\end{figure*}

\begin{figure*}[p]
\centering
\includegraphics[width=\linewidth]{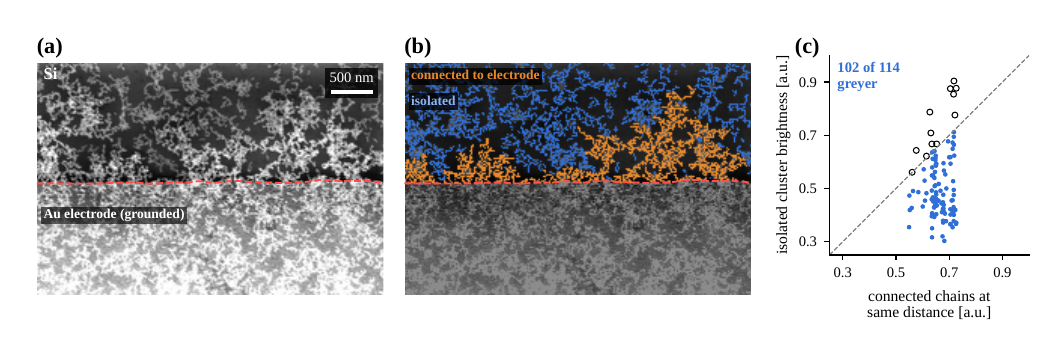}
\caption{\textbf{Second region of the same sample.}
(a)~Network at the grounded electrode (\(5\)~kV, \(0.2\)~nA, \(10\)~\(\mu\)s, \(1.35\)~nm per pixel); dashed line, foot of the electrode sidewall; display scaled to the 99.5th percentile. (b)~Connected chains (orange) and isolated clusters (blue). (c)~Brightness of each isolated cluster within \(1\)~\(\mu\)m of the electrode against connected chains at the same distance; \(102\) of \(114\) lie below the line of equal brightness.}
\label{fig:S8}
\end{figure*}

\textit{Exceptions.} Of the six main-region clusters that are not greyer (open circles in Figure~\ref{fig:S7}e), five lie within \(5\%\) of the connected reference; three of these lie about \(2\)~nm from the network, where a connection below the resolution is likely.

\textit{Limitations.} Connectivity and brightness are taken from the same image, so their agreement is not an independent electrical measurement, and gaps below about \(2\)~nm are near the resolution limit.

\FloatBarrier
\newpage

\bibliography{reference}